\documentclass[journal]{IEEEtran}
\ifCLASSINFOpdf
\else
\fi
\ifCLASSOPTIONcompsoc
 \usepackage[caption=false,font=normalsize,labelfont=sf,textfont=sf]{subfig}
\else
 \usepackage[caption=false,font=footnotesize]{subfig}
\fi
\usepackage{changepage}
\usepackage{gensymb}
\usepackage{multirow}
\usepackage{booktabs}
\usepackage{threeparttable}
\usepackage{graphics} % for pdf, bitmapped graphics files
\usepackage{epsfig} % for postscript graphics files
\usepackage{float}
\usepackage{color}
\usepackage{cite}
\usepackage{array}
\usepackage{diagbox}
\usepackage{amsmath}
\usepackage{balance}
\usepackage{wrapfig}
\usepackage{enumerate}
\usepackage{amsfonts,amssymb}
\usepackage{amsmath}
\usepackage{tabularx}
\usepackage{stfloats}
\usepackage{yhmath}
\usepackage{setspace}
\usepackage{makecell}

\usepackage{grffile}
\usepackage{mathrsfs}
\usepackage{amsmath, bm}

\usepackage[utf8]{inputenc}
\usepackage[table]{xcolor}
\usepackage{graphicx}

\usepackage[ruled, longend, linesnumbered,lined,boxed,commentsnumbered]{algorithm2e}

\begin{document}
\title{A Parameter Adaptive Trajectory Tracking and Motion Control Framework for Autonomous Vehicle}

\author{Jiarui Song,
        Yingbo Sun,
        Qing Dong,
        and Xuewu Ji
        % <-this % stops a space
\thanks{This work was supported in part by the National Natural Science Foundation of China (51975311). (Corresponding author: Xuewu Ji.)\\
\indent Jiarui Song, Yingbo Sun, Qing Dong and Xuewu Ji are with the School of Vehicle and Mobility, Tsinghua University, Beijing, 100084, China (e-mail: thusongjiarui@163.com; sun-yb20@mails.tsinghua.edu.cn; dongqing2019thu@163.com; jixw@mails.tsinghua.edu.cn).}% <-this % stops a space
}

\maketitle

\begin{abstract}
% \textcolor{red}{This paper studies the steering and braking coordination control in path tracking problems of intelligent commercial vehicles (ICVs). A gain-scheduled LPV/$H_{\infty}$ strategy for ICVs is proposed to improve tracking performance, driving stability and robustness of controller.}
This paper studies the trajectory tracking and motion control problems for autonomous vehicles (AVs). 
A parameter adaptive control framework for AVs is proposed to enhance tracking accuracy and yaw stability.
While establishing linear quadratic regulator (LQR) and three robust controllers, the control framework addresses trajectory tracking and motion control in a modular fashion, without introducing complexity into each controller.
The robust performance has been guaranteed in three robust controllers by considering the parameter uncertainties, mismatch of unmodeled subsystem as well as external disturbance, comprehensively.
Also, the dynamic characteristics of uncertain parameters are identified by Recursive Least Squares (RLS) algorithm, while the boundaries of three robust factors are determined through combining Gaussian Process Regression (GPR) and Bayesian optimization machine learning methods, reducing the conservatism of the controller.
Sufficient conditions for closed-loop stability under the diverse robust factors are provided by the Lyapunov method analytically.
The simulation results on MATLAB/Simulink and Carsim joint platform demonstrate that the proposed methodology considerably improves tracking accuracy, driving stability, and robust performance, guaranteeing the feasibility and capability of driving in extreme scenarios.
\end{abstract}
% \textcolor{red}{Sufficient conditions for closed-loop stability under the diverse robust factors are provided by the Lyapunov method analytically, which ensures the feasibility of system. 
% The results of simulations on MATLAB/Carsim platform demonstrate that the proposed methodology considerably enhance tracking accuracy, driving stability, and robust performance, which guarantees the feasibility and capability of driving in diverse scenarios.}


% \fi
\IEEEpeerreviewmaketitle
\section{Introduction}
\IEEEPARstart{R}{ecently}, with the development of vehicle technology, the automation of vehicles has gradually become one of the major developing trends \cite{1,2}. Numerous studies have shown that AVs can improve macroscopic transportation system efficiency as well as microcosmic driving safety and comfort \cite{3,Dongqing_2}. Research on AVs' autonomous vehicle technology is becoming an increasingly urgent demand and research hot-spot.

\subsection{Motivation}
% In terms of application, the relatively simple and fixed driving scenarios of commercial vehicles provide good basis for the implementation of automation, which in turn improves vehicle safety, comfort and transportation efficiency, and saves energy and labor costs \cite{3,3_}.

Trajectory tracking and motion control is one of the most important research priorities in autonomous driving technology. As highly nonlinear dynamic systems with multiple robustness factors \cite{5}, AVs place high demands on trajectory tracking controllers in terms of reconciling tracking performance with controller complexity, enabling controllers to ensure tracking accuracy, driving stability, and robustness. According to the above performance requirements, the existing control schemes are dedicated to de-complexify the controllers by designing more reasonable frameworks while improving the comprehensive tracking performance \cite{6,7}.

\subsection{Related Works}
In recent years, numerous studies on trajectory tracking and motion control of AVs have been proposed. Several works use linear control methods represented by proportional-integral-derivative (PID) \cite{8,9,10}, linear-quadratic regulator (LQR) \cite{11,12,13}, and model predictive control (MPC) \cite{duibi_MPC,14,15}, which are characterized by their simple structure and ease of implementation. The PID control considers vehicle driving state parameters such as position and heading deviation information as inputs, based on which Shi \emph{et al}. \cite{10} designes a road-curvature-range dependent PID controller tuning scheme for path tracking. Based on LQR, a discrete-time preview steering controller is proposed in \cite{13} for the servo-loop path tracking control of automated vehicles, which incorporates the time-varying disturbances over the preview window into the state vector and formed an augmented generalized linear quadratic problem. MPC predicts the future road shape, then minimizes the gap between the reference path and the trajectory predicted by the vehicle dynamics model in a receding horizon, and finally generates the optimal steering through online optimization. Cui \emph{et al}  \cite{15}  proposes a MPC-based steering angle envelope path tracking controller, in which the constraints in terms of road sides and lateral stability are directly imposed on the algorithm. 
However, the above methods are only applicable to near-linear control systems and does not apply to address the case where the state space is out of the linear conservative region. 
Some studies \cite{16,17} applied nonlinear model predictive control (NMPC) methods to nonlinear vehicle dynamics, but nonlinear optimization will introduce excessive computational complexity, and its results will be replaced by a suboptimal solution when dealing with nonlinear dynamics problems.

% However, the above methods are only applicable to near-linear control systems and cannot address the case where the state space is out of the conservative region. Some studies \cite{16,17} applied nonlinear model predictive control (NMPC) methods to nonlinear vehicle dynamics, but still can not find the optimal solution when dealing with cross-coupled dynamics problems.

% Compared with linear systems, accurate modeling of nonlinear dynamic systems of AVs can effectively improve the tracking performance. But the uncertainty of their parameters or the unstructured dynamics require the gain scheduling control to be robust under any operating conditions, which requires the design of a robust gain scheduling law \cite{18}. 

Compared with linear systems, accurate modeling of nonlinear dynamic systems of AVs can effectively improve the tracking performance.
But the multiple robust factors in nonlinear dynamics, such as parameter uncertainty, the model mismatch of unmodeled subsystem and external disturbance requires the controller to guarantee robustness under any operating conditions.
In order to address the above robustness requirements, robust controllers are applied in this field. 
By exploiting the fact that the sliding mode control (SMC) is insensitive to the uncertainty of vehicle model parameters, external disturbances and modeling errors, Guo \emph{et al}. \cite{19}  proposes an adaptive hierarchical control framework based on SMC and pseudo-inverse method. In addition, robust MPC (RMPC) \cite{20,21,22} is also frequently employed. Liang \emph{et al} \cite{20} designes a holistic adaptive multi-model predictive path tracking scheme to improve robustness of controller based on RMPC. It also becomes a mainstream technique to solve nonlinear dynamic systems with linear matrix inequalities (LMI) \cite{duibi_LMI,23,24}. In \cite{23}, an LMI-based controller is proposed which considers uncertain features such as mass, tire cornering stiffness, and vehicle velocity, which obtains the optimal solution via Lyapunov asymptotic stability. 
Dong \emph{et al} \cite{Dongqing} designs a robust control strategy for varying parameter in heavy vehicles, which ensure the lateral and roll stability of heavy vehicles by steering and braking coordination.
However, duo to a complex methodology reduces the practicality and feasibility of control system, the aforementioned methods choose to simplify vehicle dynamics and partially ignore robust factors for trade-off between complexity and performance.
Hence, a framework just with single controller is unsuitable for managing the robust control problems in complex systems.

% Sun \cite{18} \emph{et al} designes a gain-scheduled robust control strategy based on LPV/$H_{\infty}$ for varying parameter and external disturbance in heavy vehicles, which ensure the lateral and roll stability of heavy vehicles.

In addition, the boundaries of robust factors are critical parameters for a robust controller, which are typically defined as fixed maximum thresholds by the aforementioned methods to meet the system's robustness requirements. To determine the parameters of vehicle and boundary of robust factor accurately, the Least Squares or Kalman filtering techniques \cite{RLS1, RLS2, RLS_3, RLS_5} have been designed to identify the uncertain parameters in vehicle model, such as sprung mass and yaw moment of inertia. 
Nam \emph{et al} \cite{RLS_5} introduces a methodology that utilizes lateral tire force sensors to estimate the sideslip angle and tire cornering stiffness of vehicle, enhancing the control performance and driving stability of vehicles.
Sun \emph{et al} \cite{18} designs a Gaussian Process Regression (GPR) model to calculate the boundary of external disturbance, which is adopted in a gain-scheduled robust control strategy based on LPV/$H_{\infty}$, improving the robustness of controller. In addition, due to the requirements for system robustness, the robust boundary must be defined sufficiently large to guarantee the stability of system, while an excessively large robust boundary can also introduce conservatism. Therefore, a trade-off is necessary to adjust the controller's robustness parameters, thereby enhancing the system's comprehensive performance.

% Sun \cite{18} \emph{et al} designed a gain-scheduled robust control strategy based on LPV/$H_{\infty}$, in which a Gaussian Process Regression (GPR) model is adopted to calculate the boundary of external disturbance, improving the robustness of controller.

% However, the above single-controller frameworks cannot achieve the optimal control effects, which is mainly because the proposed methods ignore interference in controllers in order to optimize tracking performance and model complexity. 

% The controller framework has significant advantages in a modular fashion. Also, several mathematical and machine learning methods such as Recursive Least Squares (RLS) algorithm \cite{26}, Gaussian Process Regression (GPR) \cite{27}, and Bayesian optimization \cite{28} are incorporated into existing works to account for parameter uncertainties as well as external disturbances.

\subsection{Contribution}
Given the limitations of the above methods, a parameter adaptive trajectory tracking and motion control framework for autonomous vehicles is proposed in this paper. Distinct from the former methodologies \cite{duibi_MPC, duibi_LMI} the key contributions of this paper can be summarized as:

1) This control framework adopts LQR and robust controllers in a modular fashion, isolating the trajectory tracking problem from motion control, synchronously improving tracking performance and driving stability without introducing complexity into each controller.

2) A synthesis robust strategy based on LMI, SMC and back-stepping controller (BSC) controllers is proposed, comprehensively considering three robust factors including parameter uncertainties, mismatch of unmodeled subsystem and external disturbance. This enhances the robust performance and guarantees the asymptotic stability of the system.

3) The range  of uncertain parameters and the boundaries of robust factors are determined by the Recursive Least Squares (RLS) identification and Gaussian Process Regression (GPR) model, which are furthermore adjusted by Bayesian optimization,  improving controller accuracy and reduces the conservatism of the controller.

The rest of this paper is organized as follows: 
Section II presents preliminaries for control framework derived from vehicle-tire dynamic model and trajectory tracking error model.
Section III proposes the specific synthesis mechanism of control framework as well as the LQR and robust control strategy.
Section IV describes the parameter adaptive methodology and the determination of robust factor boundaries. 
Based on joint simulation platform, the proposed framework is verified in Section V. 
Conclusions are presents in Section VI.

\section{System models for controller design}

% \textcolor{red}{In this section, a 5-DoF simplified commercial vehicle lateral-yaw-roll dynamic model is given firstly. On this basis, the final 7-DoF lateral automated path tracking control model is deduced by introducing the lateral displacement error and yaw angle error from the path-tracking model, which will be used for the following LPV/${H_{\infty}}$ controller designing.}

In this section, for the control framework designing, a 7-DoF autonomous vehicle longitudinal-lateral-yaw dynamic model and tire slip dynamic model are first presented.
Furthermore, the trajectory tracking error model in the vehicle local reference frame is introduced, which will be utilized in the following trajectory tracking controller design.

\subsection{Motion Model of AVs}

The dynamical model of the AV is an extension of a simplified bicycle model. 
The vehicle model adopts Ackermann steering and an individual four wheel-driven system to represent the effects of lateral skidding and longitudinal slip on each tire. 
Lateral motion is also considered in the AV dynamics to include kinematic constraints of the nonholonomic system. 
Longitudinal forces are generated by tractive force exertion and friction forces to account for longitudinal slippage on each tire. 
The diagram of the AV dynamics model is presented in Fig. 1. The  longitudinal and lateral tire forces are denoted by $F_x$ and $F_y$, respectively. Inertia and force balance equations with respect to the Center of Mass (CoM) of the AVs are as follows:

\begin{figure}[h]
    \captionsetup{font={scriptsize}}
    \centerline{\includegraphics[width=0.43\textwidth]{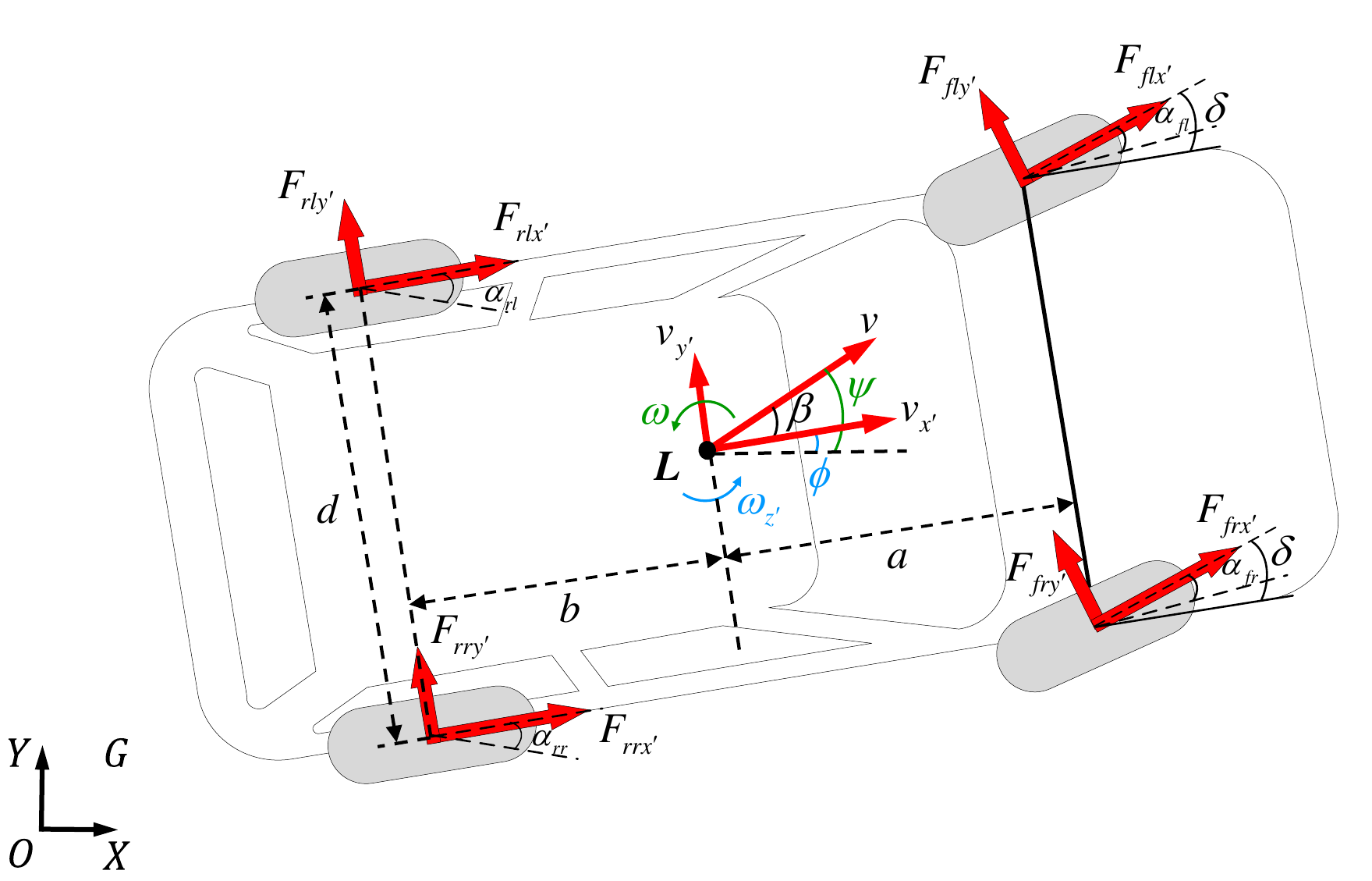}}
    \caption{Schematic representation of the autonomous vehicle model and its parameters. The modeling notation depicts forces $F_{ij}$ for each tire and vehicle motion in the local reference frame $L$.}
    \label{fig}
\end{figure}

\begin{equation}
\label{eq:vehicle_dynamic_system}
\begin{aligned}
& m\left(\dot{v}_{x^{\prime}}-v_{y^{\prime}} \omega_{z^{\prime}}\right)=F_{f r x^{\prime}}+F_{r r x^{\prime}}+F_{f l x^{\prime}}+F_{r l x^{\prime}}+F_{e x^{\prime}} \\
& m\left(\dot{v}_{y^{\prime}}+v_{x^{\prime}} \omega_{z^{\prime}}\right)=F_{f r y^{\prime}}+F_{r r y^{\prime}}+F_{f l y^{\prime}}+F_{r l y^{\prime}}+F_{e y^{\prime}} \\
& I_{z^{\prime}} \dot{\omega}_{z^{\prime}}=d\left(F_{f r x^{\prime}}+F_{r r x^{\prime}}-F_{f l x^{\prime}}-F_{r l x^{\prime}}\right)+a\left(F_{f r y^{\prime}}+F_{f l y^{\prime}}\right) \\
& ~~~~~~~~-b\left(F_{r r y^{\prime}}+F_{r l y^{\prime}}\right) 
\end{aligned}
\end{equation}
where the subscripts $\{f, r\}$ stand for front or rear tires, and $\{l, r\}$ for left or right tires, respectively. The AV chassis is described as a rigid body represented by its position in the geometric center, linear speeds of the vehicle $v_{x'}$, $v_{y'}$ and angular speeds $\omega_{z'}$ in the local reference frame $L$, as shown in Fig. 1. In addition, the tire dynamics, through force and inertia balance on each wheel-motor, are described by:

\begin{equation}
\label{eq:wheel_system}
J_\omega \dot{\omega}_{i j}+B_e \omega_{i j}=T_{i j}+T_f-r_\omega F_{i j x^{\prime}}
\end{equation}
where $i \in\{f, r\}$ and $j \in\{l, r\}$ denote the location of a tire in the vehicle; $r_{\omega}$ is the effective tire radius; $J_\omega$ and $B_e$ denote the moment of inertia and damping coefficient of tire, respectively; $T_{f}$ is the frictional resistance moment; $w_{ij}$ denote the angular speed for each tire; $T_{ij}$ is the motor torque applied to each wheel axle.

\subsection{Model of the Tire Slip Dynamics}

The slip ratio $\sigma$ and side-slip angle $\alpha$ provide an idea about the tire mechanics and methods for calculating tire forces. The slip ratio $\sigma$ for each tire is defined by:

\begin{equation}
\label{eq:slip_function}
\sigma_{ i j}=\left\{\begin{array}{l}
\frac{s_{i j}}{\omega_{i j} r_{\omega}}, \text { if } \omega_{i j} r>v_{i j x^{\prime}}, \omega_{i j} r \neq 0, \text { for driving, } \\
\frac{s_{i j}}{v_{i j x^{\prime}}}, \text { if } \omega_{i j} r<v_{i j x^{\prime}}, v_{i j x^{\prime}} \neq 0, \text { for braking. }
\end{array}\right.
\end{equation}
where $s_{i j}=\omega_{i j} r_{\omega}-v_{i j x^{\prime}}$, $s_{ij}$ denotes the relative linear speed of $\omega_{ij} r_{\omega}$ for each tire with respect to $v_{ijx'}$ on the longitudinal axis of the vehicle.

The tire side-slip angle $\alpha$ represents the angle between the tire velocity $v_{ij'}$ and the longitudinal axis of the tire. Since in AVs the lateral velocities of tires on the same transverse axis are very similar, each pair of lateral tires results in similar tire skidding. Thus, to reduce the model complexity, it is presumed that each pair of lateral tires experience the same side-slip angle, $\alpha_{fr} = \alpha_{fl}$, $\alpha_{rr} = \alpha_{rl}$. The tire side-slip angles will be approximately represented by:

\begin{equation}
\begin{aligned}
    \alpha_{fj} = \frac{v_{x'}+a\omega_{z'}}{v_{y'}} - \delta, ~~ \alpha_{rj} = \frac{v_{x'}-b\omega_{z'}}{v_{y'}}.
\end{aligned}
\end{equation}
where $\delta$ denotes the steering angle. The longitudinal $F_{ijx'}$ and lateral $F_{ijy'}$ tire forces depend essentially on the vertical load $F_{ijz'}$ along with the slip ratio $\sigma_{ij}$ and side-slip angle $\alpha_{ij}$, which describe the most nonlinear behavior of the friction forces given the complex tire–terrain relationship. Then, the Dugoff tire force nonlinear model is adapted here to capture such nonlinearities. Longitudinal and lateral tire forces are formulated as:

\begin{equation}
\begin{aligned}
    &F_{ijx'} = \frac{C_{\sigma}\sigma_{ij}}{1 + \sigma_{ij}}f\left( \lambda_{ij} \right),~~  F_{ijy'} = \frac{C_{\alpha}{ \tan\alpha}_{ij}}{1 + \sigma_{ij}}f\left( \lambda_{ij} \right).\\
    &\lambda_{ij} = \frac{\mu F_{ijz'}\left( 1 + \sigma_{ij} \right)}{2\sqrt{{\left( C_{\sigma}\sigma_{ij} \right)^{2} + \left( C_{\alpha}{ \tan\alpha}_{ij} \right)}^{2}}}
\end{aligned}
\end{equation}
where $C_{\sigma}$ and $C_{\alpha}$ denote the longitudinal and lateral tire stiffness, respectively;  $F_{ijz'}$ is the unvarying and uniformly distributed vertical force across the vehicle chassis; $\mu$ denotes the ground adhesion coefficient; $\lambda_{ij}$ denotes attachment reserve coefficient.

\subsection{Trajectory Tracking Error Model}

In the interest of notational simplicity, it has been omitted the expression of the local coordinate frame $L$ in relation to the dynamics of the vehicle speeds. This means that we assume the vehicle speeds are equivalent to the local coordinate frame $L$, i.e., $v_x = v_x', v_y = v_y', v_z = v_z', \omega_y = \omega_y', \omega_z = \omega_z'$. Furthermore, the equations representing the vehicle kinematics in $L$ with respect to the global reference frame $G$ are as follows:

\begin{equation}
\begin{aligned}
\dot{x} & =v \cos \psi \\
\dot{y} & =v \sin \psi \\
\dot{\psi} & =\omega = \omega_z + \dot\beta
\end{aligned}
\end{equation}
where $x$ and $y$ represent the global position of the vehicle fixed at $L$; $\psi$ denotes the yaw angle of the vehicle;  $\omega$ denotes the yaw rate. $\omega_z$ denotes the vertical angular speed of the vehicle; $\beta$ is the side slip angle of CoM. The vehicle speed $v$ represents  the combination of the longitudinal velocity $v_x$ and lateral velocity $v_y$ in the reference frame $G$.

% The kinematics components of the vehicle speed $v$ are described by the longitudinal $v_x$ and lateral $v_y$ in the reference frame $G$.

As one of the control problems focuses on tracking a desired trajectory, a trajectory tracking error model is raised  to account for the vehicle pose and kinematics. To establish this model, tracking errors are determined by the difference between the reference trajectory in global frame $G$ and the global vehicle states, which are subsequently mapped into the vehicle local reference frame attached to $L$ as follows:

\begin{equation}
\bm{z_e}=\left[\begin{array}{ccc}
\cos \psi & \sin \psi & 0 \\
-\sin \psi & \cos \psi & 0 \\
0 & 0 & 1  \\
\end{array}\right](\bm{z} - \bm{z}^{\text {ref}})
\end{equation}
where the global vehicle states $\bm{z} = [x, y, \psi]^T$, the vector of reference trajectory $\bm{z}^{\text {ref}} = [x^{\text {ref}}, y^{\text {ref}}, \psi^{\text {ref}}]^T$, the trajectory tracking error $\bm{z_e} = [e_x, e_y, e_{\psi}]^T$. The model of the trajectory tracking error dynamics is given by:

\begin{equation}
\label{eq:tracking_model}
\begin{aligned}
\dot{e}_x & =v - v^{\text {ref }} \cos \left(e_\psi\right)+e_y \omega_z \\
\dot{e}_y & =v^{\text {ref }} \sin \left(e_\psi\right)-e_x \omega_z \\
\dot{e}_\psi & =\omega - \omega^{\text {ref }}
\end{aligned}
\end{equation}

The tracking error model in \eqref{eq:tracking_model} is linearized around reference points: $\bm{z} = \bm{z}^{\text{ref}}$, $\bm{u} = \bm{u}^{\text{ref}}$, and reference control input $\bm{u}^{\text{ref}} = [v^{\text{ref}} ~~ \omega^{\text{ref}}]^T$. Thus, the tracking error model becomes:

\begin{equation}
\label{eq:linear_tracking_model}
\dot{\bm z}_{\bm e}(t)=\bm{A_e}(t) \bm{z_e}(t)+\bm{B_e}(t) \bm{u_e}(t)
\end{equation}
where the elements of the matrix $\bm{A_e}$ and matrix $\bm{B_e}$ are given by:

\begin{equation}
    \bm{A_e} = \begin{bmatrix}
        0 & \omega^{\text {ref}} & 0 \\
        -\omega^{\text {ref}} & 0 & v^{\text {ref}} \\
        0 & 0 & 0 \\
        \end{bmatrix}~\bm{B_e} = \begin{bmatrix}
        1 & 0 \\
        0 & 0 \\
        0 & 1 \\
        \end{bmatrix}.
\end{equation}
where $\bm{{A}_e}(t)$ and $\bm{{B}_e}(t)$ are matrices of the linear error model and the control input vector $\bm{{u}_e}(t) = [e_v ~~ e_{\omega}]^T$.The controllability matrix \( \bm{{C}_o} = [ \bm{{B}_e} \quad  \bm{{A}_e} \bm{{B}_e} \quad  \bm{{A}_e}^2 \bm{{B}_e} ]\) associated to the error model \eqref{eq:linear_tracking_model} has full rank if the reference speed or reference yaw rate is nonzero. Thus, all the tracking errors can be forced to zero by a full state feedback, when all reference values are avoided being zero simultaneously.

\section{Controller design}

The overall structure of the control framework is depicted in Fig. 2, in which controllers are orchestrated in a modular fashion.
LQR controller, designed to address trajectory tracking problem, utilizes a kinematic model in vehicle local reference frame and calculates a reference motion for the longitudinal-lateral-yaw dynamics.
Building on this foundation, three robust controllers, LMI, SMC and BSC, are tailored to track the calculated reference dynamic motion while comprehensively managing various robust factors, such as parameter uncertainties, mismatch of unmodeled subsystem and external disturbance.
Additionally, a parameter adaptive strategy is integrated into this framework to determine and adjust both the range of uncertain parameters and the boundaries of robust factors, which will be expounded in Section IV.

\begin{figure}[htbp]
    \captionsetup{font={scriptsize}}
    \centerline{\includegraphics[width=0.5\textwidth]{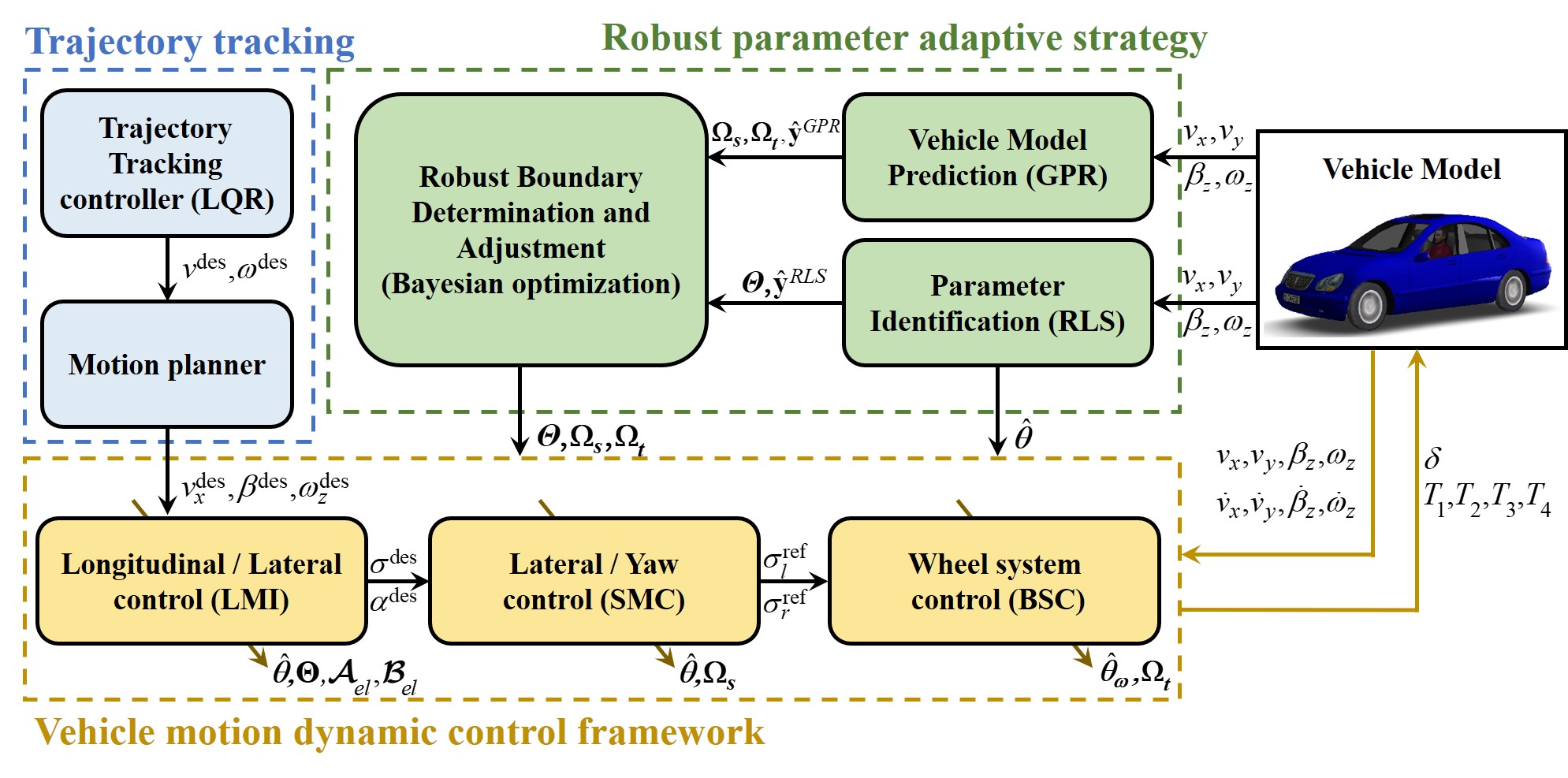}}
    \caption{Illustration of the parameter adaptive trajectory tracking and motion control framework: the LQR controller and motion planner address trajectory tracking problem and calculate the 3D phase trajectory, while three robust controllers deal with the motion control problem. LMI controller manages the longitudinal-lateral dynamics and calculates the desired slip ratio and tire side-slip angle; SMC controller manages the lateral-yaw dynamics and addresses yaw stability controller problems; BSC controller manages wheel system. The parameters in these robust controllers are adjusted by parameter adaptive strategy. RLS identifies the uncertain parameters of vehicle; GPR characterize the vehicle dynamics and calculates the boundary of model mismatch and external disturbance; Bayesian optimization determines and adjusts the above parameters which will be adopted in robust controllers.}
    \label{fig}
\end{figure}

\subsection{LQR-based Trajectory Tracking Controller Design}\label{AA}

% \begin{equation}
% \dot{\bm z}_{\bm  e}(k+1)=\bm{A_k} \bm{z_e}(t)+\bm{B_k} \bm{u_e}(k)
% \end{equation}
% where $\bm{A_k} = \bm{I} + T\bm{A_e}\text{,}\ \bm{B_k} = T\bm{B_e}$, and $\bm{I}$ denote the appropriate dimension matrices; $T$ is the time step of the MPC controller.

% \begin{equation}
%     \begin{gathered}
%         \min _{\bm{u_e}} J(k)=\sum_{i=1}^{\infty}\left\|\bm{z_e}(k+i )\right\|_{\bm{Q_k}}^{2}\!\!\!\!~+~\!\!\sum_{i=0}^{\infty}\| \bm{u_e}\left( k+i \right )\|_{\bm {R_k}}^{2}
%     \end{gathered}
% \end{equation}

As the trajectory tracking error model given in Eq. (6)–(9), the LQR controller can export a control variable to minimize the tracking error by solving the quadratic programming problem given by:
% \begin{equation}
%     \begin{gathered}
%         \min _{\Delta\widetilde{ \bm{\mathcal{U}}}, \varepsilon} J(k)=\sum_{i=1}^{N_{p}}\left\|\widetilde{\bm{\mathcal{X}}}(k+i )\right\|_{\bm{Q_m}}^{2}\!\!\!\!+\!\!\sum_{i=0}^{N_{c}-1}\| \Delta\widetilde{\bm{\mathcal{U}}}\left( k+i \right )\|_{\bm R_m}^{2}
%     \end{gathered}
% \end{equation}

\begin{equation}
\label{eq:trakcing_performace}
J=\int_0^{\infty}\left(\left\|\bm{z_e}(t)\right\|_{\bm{Q_k}}^{2}+\|\bm{u_e}\left(t \right)\|_{\bm {R_k}}^{2}\right) d t
\end{equation}
where $\bm {Q_k}$ and $\bm {R_k}$ are the weight coefficient matrices. As a control objective for trajectory tracking, the controller is required to ensure asymptotic stability of the system under optimal performance \eqref{eq:trakcing_performace}. Therefore, the control input of the tracking error system \eqref{eq:linear_tracking_model} is designed based on the LQR state-feedback control law:
\begin{equation}
    \bm{u_e}(t)=-\bm {K_k}\bm{z_e}(t)
\end{equation}
where the feedback gain $\bm{K_k}$ can be calculated by:
\begin{equation}
    \bm {K_k}=\bm{R_k}^{-1}\bm{B_e}^{T}\bm{P_k}
\end{equation}
where $\bm{P_k}$ is solved from the Riccati equation as follows:

\begin{equation}
\label{eq:Riccati}
    \bm{A_e}^{T}\bm{P_k}+\bm{P_k}\bm{A_e}+\bm{Q_k}-\bm{P_k}\bm{B_e}\bm{R_k}^{-1}\bm{B_e}^{T}\bm{P_k}=0
\end{equation}
where the tracking error system is asymptotically stable if there exists a positive definite symmetric matrix $\bm {P_k} = \bm {P_k}^T > 0$, obtained by solving Riccati equation \eqref{eq:Riccati}. The desired control input is calculated by:

\begin{equation}
    \bm u(t)=  \bm{u_e}(t) + \bm u^{\text{ref}}
\end{equation}
where $\bm u(t)=[v^{\text{des}}, \omega^{\text{des}}]^{\mathrm{\rm T}}$ consisting of the desired vehicle speed $v^{\text{des}}$ and the desired yaw rate $\omega^{\text{des}}$.

% \textcolor{red}{-----------------------------------------------------}

By calculating the derivative of side slip angle of CoM, the tire adhesion margin is optimized to enhance the stability of vehicle motion, while the yaw rate can be allocated to the side slip angle of CoM ${\beta}$ and vertical angular speed $\omega_z$. The optimization objective function containing the rate of tire friction utilization ${\varphi}_i$ is defined by:
% \begin{equation}   
% \begin{aligned}
% \varphi_{i}=\frac{F_{x i}^{2}+F_{y i}^{2}}{\left(\mu F_{z i}\right)^{2}} \triangleq \varphi_{i}(\dot{\beta}^{\text{des}})\\
% \end{aligned}
% \end{equation}

% \begin{equation}   
% \begin{aligned}
% &\min _{\dot{\beta}^{\text{des}}} J_{\beta}(t)=\sum_{i=1}^{4}{\varphi_{i}} + W_{\beta}({\dot{\beta}^{\text{des}}})^2 \\
% &~~~~~~\text { s.t. } 0 \leq \varphi_{i} \leq 1\\
% \end{aligned}
% \end{equation}

\begin{equation}   
\label{eq:optimiazation}
\begin{aligned}
&\min _{\dot{\beta}^{\text{des}}} J_{\beta}(t)=\sum_{i=1}^{4}{\varphi_{i}} + W_{\beta}{\dot{\beta}}^2 \\
&~\text { s.t. } \varphi_{i}(\dot{\beta})=\frac{F_{x i}^{2}+F_{y i}^{2}}{\left(\mu F_{z i}\right)^{2}}\\
&~~~~~~~~~~ 0 \leq \varphi_{i} \leq 1\\ 
\end{aligned}
\end{equation}
where $W_{\beta}$ is smoothing factor of  $\dot{\beta}$, while $\dot{\beta}^{\text{des}}$ can be calculated by solving the nonlinear optimization problem \eqref{eq:optimiazation}. If the computational efficiency is insufficient, based on the small angle assumption $ {\rm tan}{\alpha_i} \approx {\alpha_i}$, $\varphi_{i}$ and $J_{\beta}$ can be reduced to a quadratic function of $\dot{\beta}$.

Furthermore, the desired longitudinal speed $v_{x}^{\text{des}}$, desired side slip angular of CoM ${\beta}^{\text{des}}$ and desired vertical angular speed $\omega_z^{\text{des}}$ of vehicle can be calculated as:
\begin{equation}   
\begin{aligned}
{\beta}^{\text{des}}(t)&={\beta}^{\text{des}}(t-T)+\dot{\beta}^{\text{des}}(t)T\\
v_{x}^{\text{des}}(t) &= v^{\text{des}}(t) {\rm cos} {\beta}^{\text{des}}(t)\\
\omega_z^{\text{des}}(t)&=\omega^{\text{des}}(t)-\dot{\beta}^{\text{des}}(t)
\end{aligned}
\end{equation}
where $T$ represents the time step of the controller. The derivative of longitudinal speed ${\overset{.}{v}}_{x}^{\text{des}}$ and vertical angular speed  $\dot{\omega}_z^{\text{des}}$ can be calculated by using a discrete differencing method.

\begin{equation}
\begin{aligned}
    {\overset{.}{v}}_{x}^{\text{des}}(t) &= \left[ v_{x}^{\text{des}}\left( k \right) - v_{x}^{\text{des}}\left( t-T \right) \right]/T \\  
    \dot{\omega}_z^{\text{des}}(t) &=[\omega_z^{\text{des}}(t) - \omega_z^{\text{des}}(t-T)]/T 
\end{aligned} 
\end{equation}

The 3D phase trajectory $\bm{T_{rj}}=\left\lbrack v_{x}^{\text{des}},\beta^{\text{des}},\omega_z^{\text{des}} \right\rbrack^{\rm T}$  will be tracked by three robust controllers with the reformulated vehicle dynamic system (19), which are explained in detail in the following.

\begin{equation}
\label{eq:vehicle_model}
\begin{aligned}
 \dot{v}_{x}=&\frac{1}{m}(F_{f r x}+F_{r r x}+F_{f l x}+F_{r l x}+F_{e x}) + v_{y} \omega_{z} \\
 \dot{\beta}_{z} =&\frac{1}{m{v}_{x}}(F_{f r y}+F_{r r y}+F_{f l y}+F_{r l y}+F_{e y}) - \omega_{z} \\
 \dot{\omega}_{z}=&\frac{1}{I_{z}}[d\left(F_{f r x}+F_{r r x}-F_{f l x}-F_{r l x}\right)+a\left(F_{f r y}+F_{f l y}\right) \\
& -b\left(F_{r r y}+F_{r l y}\right)]
\end{aligned}
\end{equation}

% \begin{equation}
% \begin{gathered}
% \min _{\dot{\beta}^{\text {des }}} J_\beta(t)=\sum_{i=1}^4 \varphi_i+W_\beta\left(\dot{\beta}^{\text {des }}\right)^2 \\
% \text { s.t. } 0 \leq \varphi_i \leq 1 \\
% \varphi_i(\dot{\beta})=\frac{F_{x i}^2+F_{y i}^2}{\left(\mu F_{z i}\right)^2}
% \end{gathered}
% \end{equation}

% \textcolor{red}{-----------------------------------------------------}

\subsection{LMI-based Vehicle Longitudinal-Lateral Dynamic Robust Controller Design}\label{AA}

% \textcolor{red}{Since the trajectory tracking accuracy is mainly related to the longitudinal and yaw motions of the vehicle, the nonlinear system, composed of the two coupled longitudinal and yaw dynamics systems in (1), is reformulated as follows:}

% \begin{equation}
% \begin{aligned}
%     \overset{.}v_x &= \frac{1}{m}(F_{f r x^{\prime}}+F_{r r x^{\prime}}+F_{f l x^{\prime}}+F_{r l x^{\prime}}+F_{e x^{\prime}})+v_yr,\\
%     &\triangleq f_{x}(\bm{x_l}, \bm{u_l}),\\
%     % \overset{.}v_y&= \frac{1}{m}({F}_{xf}\sin \delta  - {F}_{yf}\cos\delta + {F}_{yr} - F_{y \rm dist} - mv_xr), \\
%     % &\triangleq f_{1}(\bm{x_s}, u_s),\\
%     \dot{r}&= \frac{1}{{I_{z}}}(l_{ f} F_{x f} \sin \delta+l_{f} F_{y f} \cos \delta-l_{ r} F_{y r})\triangleq f_{r}(\bm{x_l}, \bm{u_l}).
% \end{aligned}
% \end{equation}

Since the vehicle dynamic model depends on characteristics for each type of tire, the model can be made adaptive to changes in tire stiffness by identifying the tires longitudinal and lateral stiffness parameters $\bm \theta = [C_{\sigma};C_{\alpha}]$. The nonlinear system composed of the two coupled longitudinal and yaw dynamics systems in (19) is reformulated as follows:

\begin{equation}
\dot{\bm{x_{l}}}(t)=f(\bm{x_{l}}(t), \bm u(t), \theta)
\end{equation}
where $t$ represents continuous time; $\bm{x_{l}} = \left\lbrack v_x,\omega_z \right\rbrack^{\rm T}$ and $\bm{u_{l}} = \left\lbrack {\sigma,\alpha} \right\rbrack^{\rm T}$ represent the vector of system states and control input, respectively. Furthermore, the longitudinal-yaw dynamic system \eqref{eq:vehicle_model} is linearized as an nominal error discrete system \eqref{eq:discrete_system} around reference points $\overset{.}{\bm x}_{\bm l}^{\text {ref}} = f\left( \bm{x_{l}}^{\text {ref}},\bm{u_{l}}^{\text {ref}}\right)$  as follows:

% $\overset{.}{\bm x}_{\bm l}^{\text {ref}}$
\begin{equation}
\label{eq:discrete_system}
    {\widetilde{\bm x_l}}(k+1) = \bm A\widetilde{\bm{x_l}}(k) + \bm B\widetilde{\bm{u_l}}(k),
\end{equation}
where $\widetilde{\bm{x_l}}=\bm{x_l}-\bm{x_{l}}^{\text {ref}}$ and  $\widetilde{\bm{u_l}}=\bm{u_l}-\bm{u_{l}}^{\text {ref}}$, the system matrix $\bm A$ and input matrix $\bm B$ with parameters ${\theta}$ are defined as:

\begin{equation}
\bm A=\bm{I}+T \frac{\partial f\left(\bm{x_{l}}, \bm{u_{l}}, {\bm \theta} \right)}{\partial \bm{x_{l}}},  \bm B=T \frac{\partial f\left(\bm{x_{l}}, \bm{u_{l}}, {\bm \theta}\right)}{\partial \bm{u_{l}}}
\end{equation}

% \begin{equation}
% \bm A=\bm{I}+\left.T \frac{\partial f\left(\bm{x_{l}}, \bm{u_{l}}, \theta \right)}{\partial \bm{x_{l}}}\right|_{\substack{\bm{x_{l\rm ref}} \\ \bm{y_{l\rm ref}}}},  \bm B=\left.T \frac{\partial f\left(\bm{x_{l}}, \bm{u_{l}}, \theta\right)}{\partial \bm{u_{l}}}\right|_{\substack{\bm{x_{l\rm ref}} \\ \bm{u_{l\rm ref}}}}
% \end{equation}

Given the presence of model uncertainties and identification errors, there still exists mismatch between the actual model and the nominal model based on parameter identification. For correcting the model mismatch-based errors, an uncertain error linear system, including nominal matrix $\bm{\hat{A}}, \bm{\hat{B}}$ and error matrix $\bm{{\Delta A}}, \bm{\Delta {B}}$, is designed as follows:

\begin{equation}
\label{eq:error_model}
    \overset{.}{\bm{\widetilde{x_l}}}(k+1) = \left( {\bm{\hat{A}} + \bm{\Delta A}} \right)\bm{\widetilde{x_l}}(k) + \left( {\bm{\hat{B}} + \bm{\Delta B}} \right)\bm{\widetilde{u_l}}(k).
\end{equation}
where $\bm{\hat{A}}(\hat{C}_{\alpha},\hat{C}_{\sigma})$ and $\bm{\hat{B}}(\hat{C}_{\alpha},\hat{C}_{\sigma})$ is defined by the nominal parameter vector $\bm {\hat{\theta}}$, while $n_\theta$-elements $C_{\sigma}$ and $C_{\alpha}$ of actual parameters $\bm \theta$ are specified within a bounded reference range $C_{\alpha} \in [C_{\alpha}^{min} ,C_{\alpha}^{max}]$, $C_{\sigma} \in [C_{\sigma}^{min} ,C_{\sigma}^{max}]$ at each time step. Therefore, for any permissible parameter vector $\bm \theta$ confined to a polytope $\Theta \subseteq \mathbb{R}^{n_\theta}$, the error system matrix $\bm{\Delta A} \in \mathcal{A}_{el}$ and $\bm{\Delta B} \in \mathcal{B}_{el}$, also remain bounded and defined within the polytopes $\mathcal{A}_{el}$ and $\mathcal{B}_{el}$. The parametric error model is represented by a collection of local linear systems, with each member of the collection corresponding to a vertex system. Each vertex is denoted by the $p$-th collection of matrices $\left\{\bm{A}_{el}^p, \bm{B}_{el}^p\right\}$, formed by the extreme values of the parameter vector range. Thus, the polytopic collection of the error matrix satisfy $\mathcal{A}_{el} = \text{Convh} \left\{\bm{A}_{el}^1, \bm{A}_{el}^2, \bm{A}_{el}^3, \bm{A}_{el}^4\right\}$ and $\mathcal{B}_{el} = \text{Convh} \left\{\bm{B}_{el}^1, \bm{B}_{el}^2, \bm{B}_{el}^3, \bm{B}_{el}^4\right\}$, while the error matrices $\bm{\Delta A}$ and $\bm{\Delta B}$ are defined as follows:

\begin{equation}
    \left\lbrack \bm{\Delta A}~~~\bm{\Delta B} \right\rbrack = \bm{M}\bm{F}\left( t \right)\left\lbrack \bm{N_{a}}~~\bm{{N}_{b}} \right\rbrack.
\end{equation}
where $\bm{M} = \left\lbrack {\bm{I}~~\bm{I}~~\bm{I}~~\bm{I}} \right\rbrack$, $\bm{N_{a}} = \left\lbrack \bm{A}_{el}^1, \bm{A}_{el}^2, \bm{A}_{el}^3, \bm{A}_{el}^4 \right\rbrack^{\rm T}$, $\bm{N_{b}} = \left\lbrack\bm{B}_{el}^1, \bm{B}_{el}^2, \bm{B}_{el}^3, \bm{B}_{el}^4 \right\rbrack^{\rm T}$, $\bm{I}$ is the identity matrix with appropriate dimensions; $\bm F(t)$ denotes an bounded uncertainty matrix, with the relationship $\bm F^{\rm T}\left( t \right)\bm {F}\left( t \right) \leq 1$.

As a robust control objective, the controller is required to guarantee the asymptotic stability of the error system in the presence of model uncertainties and identification errors. Hence, the control input $\bm{\widetilde{u_l}}(k)$ for the error system in \eqref{eq:error_model} is calculated based on the state-feedback control law:

\begin{equation}
\begin{aligned}
    &{\bm{\widetilde{u_l}}}\left( k \right) = \bm K{\bm{\widetilde{x_l}}}\left( k \right) \\
    &\bm{u_l}(k) = {\bm{\widetilde{u_l}}}\left( k \right) + \bm{u_{l}}^{\text {ref}}
\end{aligned}
\end{equation}
where matrix $\bm K$ is the control gain for mismatch system that keeps the states of the error dynamics as close as possible to a uncertainty-free state. Designing a quadratic Lyapunov function formed by $V(\bm{\widetilde{x_l}}) = \|\bm{\widetilde{x_l}}|^2_{\bm P^{-1}}$, the mismatch system is asymptotically stable if there exists a positive definite symmetric matrix $\bm P = \bm P^T > 0$ such that $V(\bm{\widetilde{x_l}}(k+1)) - V(\bm{\widetilde{x_l}}(k)) < 0$ for all $\Delta z_e \neq 0$. However, in order to determine the matrix $\bm P$ so that the control input ${\bm{\widetilde{u_l}}}\left( k \right)$ minimizes the robust performance cost, the stability condition for the error system in \eqref{eq:error_model} is defined  as follows:

\begin{equation}
\label{eq:stable_condition}
V(\bm{\widetilde{x_l}}(0)) \geq {\sum\limits_{k = 0}^{\infty}\left( \left\|\widetilde{\bm{{x_l}}}(k)\right\|_{\bm{Q}}^{2} + \left\|\widetilde{\bm{{u_l}}}(k)\right\|_{\bm{R}}^{2} \right)}  
\end{equation}
where $\bm Q$ and $\bm R$ are positive definite weight matrices. Thus, the stabilizing condition, considering the candidate function $V(\bm{\widetilde{x_l}}(k))$ and the condition \eqref{eq:stable_condition}, is defined by:

\begin{equation}
V(\bm{\widetilde{x_l}}(k)) - V(\bm{\widetilde{x_l}}(k+1)) \geq  \left\|\widetilde{\bm{{x_l}}}(k)\right\|_{\bm{Q}}^{2} + \left\|\widetilde{\bm{{u_l}}}(k)\right\|_{\bm{R}}^{2}
\end{equation}
Through substituting the uncertain system \eqref{eq:error_model} into the stabilizing condition \eqref{eq:stable_condition}, the new requirement can be formulated for the mismatch system as follows:

\begin{equation}
\label{eq:LMI_condition}
    \begin{aligned}
       &\left[ {\bm A + \bm{BK} + \bm{MF}\left( k \right)\left. \left( \bm{N_{a}} + \bm {N_b}\bm{K} \right. \right)} \right]^{\rm T}\bm{P}^{- 1}[ \bm A + \bm{BK} +\\
       &\bm{MF}\left( k \right)\left( \bm{N_{a}} + \bm {N_b}\bm{K} \right.)] - \bm{P}^{- 1} + \bm{Q} + \bm{K}^{\rm T}\bm{RK} < 0.
    \end{aligned} 
\end{equation}
Applying the Schur complement theorem to convert the nonlinear condition \eqref{eq:LMI_condition} into linear matrix inequalities, the stabilizing condition can be derived as:

\begin{equation}
\label{eq:LMI}
\begin{aligned}
    \begin{bmatrix}
        {- \bm{P} + \varepsilon \bm{DD}^{\rm T}}&*&0&0&0 \\
        \left( \bm{AP} + \bm B\bm{Y})^{\rm T} \right.&{- \bm{P}}&*&\bm{P}&\bm{Y}^{\rm T} \\
        0&\left. \left( \bm{N_{a}}\bm{P} + \bm {N_b}\bm{Y} \right. \right)&{- \varepsilon \bm{I}}&0&0 \\
        0&\bm{P}&0&{- \bm{Q}^{- 1}}&0 \\
        0&\bm{Y}&0&0&{- \bm{R}^{- 1}} \\
    \end{bmatrix}& \\
    < 0.&
\end{aligned}
\end{equation}

Therefore, if linear matrix inequalities \eqref{eq:LMI} can be solved for a positive definite symmetric matrix $\bm{P}$, matrix $\bm{Y}$, and positive number $\varepsilon$, the mismatch error system is asymptotically stable while the robust performance is bounded. The feedback control law can be calculated by $\bm K = \bm Y \bm P^{-1}$. The desired control input is represented  as $\bm{u_l}=[\sigma^{\text{des}}, \alpha^{\text{des}}]$, assuming equal slip ratios on both sides of the wheels, i.e., $\sigma^{\text{des}}=\sigma_l^{\text{des}}=\sigma_r^{\text{des}}$, which will be modified and reassigned by the SMC controller. The steering angle exported to the vehicle can be calculated by $\delta=\frac{v_{y}+\omega_z \times l_{f}}{v_{x}}-\alpha_{\mathrm{des}}$.

\subsection{SMC-based Vehicle Stability Robust Controller Design}\label{AA}

% In the SMC robust controller, a vehicle lateral and yaw dynamics are given by:

% \begin{equation}
% \begin{aligned}
%     \overset{.}{\beta}&= \frac{1}{mv_x}({F}_{xf}\sin \delta  + {F}_{yf}\cos\delta + {F}_{yr} - F_{y \rm dist}) - r, \\
%     &\triangleq f_1\left( \bm{x_s} \right) + g_1\left( \bm{x_s} \right)u_s,\\
%     \dot{r}&= \frac{1}{{I_{z}}}(l_{ f} F_{x f} \sin \delta+l_{f} F_{y f} \cos \delta-l_{ r} F_{y r}+ \Delta M_{z} - M_{z \rm rist})\\
%     %  &~~~~- M_{z \rm rist}),\\
%     &\triangleq f_2\left( \bm{x_s} \right) + g_2\left( \bm{x_s} \right)u_s.
% \end{aligned}
% \end{equation}

Since the vehicle stability is mainly related to lateral and yaw dynamics, which not only depend on tire stiffness parameters but also influenced by disturbance forces and moments, the nonlinear lateral-yaw dynamic system based on the CoM side-slip angle $\beta$ and angular speeds $\omega_z$ is formulated as follows:

\begin{equation}
\label{eq:SMC_system}
\overset{.}{\bm{x_s}}(t)=
f_s\left( \bm{x_s}(t), \bm{\theta} \right) + g_s\left( \bm{x_s}(t), \bm{\theta} \right)u_s + \bm{\omega_s}(t)
\end{equation}
where $\bm{x_s} = \left\lbrack \beta,\omega_z \right\rbrack^{\rm T}$ and $u_s =u_s(  {\frac{1}{1 + \sigma_{l}},\frac{1}{1 + \sigma_{r}}} )$; $\sigma_{l}$ and $\sigma_{r}$ denote the slip ratios of wheels on the left and right sides; $\bm{\omega_s}(t)$  denotes the disturbance lateral forces and yaw moments bounded in a compact set $\bm{\Omega_s} \subseteq \mathbb{R}^{n_\omega}$.

In the meanwhile, the identification errors of tire stiffness parameters and the presence of unmodeled systems will cause a mismatch between the nominal and actual models. For correcting model mismatch-based errors and restricting the impact of disturbances, a sliding mode system with uncertainty in the $\beta-\omega_z$ phase plane is designed to tracking the desired CoM side-slip angle $\beta^{\text {des}}$ and angular speeds $\omega_z^{\text {des}}$ as follows:

\begin{equation}
    \begin{aligned} 
    s &= \omega_z - \omega_z^{\text {des}} + \xi\left( \beta - \beta^{\text {des}} \right)\\
    \overset{.}{s} &= \overset{.}{\omega}_z - {\overset{.}{\omega}}_z^{\text {des}} + \xi( \overset{.}{\beta} - {\overset{.}{\beta}}^{\text {des}} )
    \end{aligned} 
\end{equation}
where $s$ denotes the sliding surface designed in the form of a hyperplane for the continuous system \eqref{eq:SMC_system}. It guides the evolution of the system states on this hyperplane according to the state trajectories specified by the sliding manifold $\omega_z - \omega_z^{\text {des}} = -\xi\left( \beta - \beta^{\text {des}} \right)$. Furthermore a nominal model based associated with the sliding surface is reformulated as follows:

\begin{equation}
\overset{.}{s} = \hat{h}\left( \bm{x_s}, \bm{\hat{\theta}} \right) + \hat{k}\left( \bm{x_s}, \bm{\hat{\theta}} \right)u_s
\end{equation}
where $\hat {h}\left( \bm{x_s} \right)$ and $\hat{k}\left( \bm{x_s} \right)$ represent the nominal system function. The actual sliding mode system model with parameter uncertainties and external disturbances is reformulated as follows:

\begin{equation}
\begin{aligned}
    \overset{.}{s} = h\left( \bm{x_s}(t), \bm{\theta} \right) + k\left( \bm{x_s}(t), \bm{\theta} \right)u_s + \omega_l(t).
\end{aligned}
\end{equation}
where $h\left( \bm{x_s} \right)$ and $k\left( \bm{x_s} \right)$ represent the actual system function, $\omega_l(t)$ denotes the coupled disturbance bounded in a compact set $\bm{\Omega_l} \subseteq \mathbb{R}$. Meanwhile the differences between the actual and nominal system in bounded form satisfy:

\begin{equation}
\label{eq:smc_error}
\begin{aligned}
&{\delta_s}\left( \bm{x_s} \right) = h\left( \bm{x_s} \right) - \hat{h}\left( \bm{x_s} \right)\frac{k\left( \bm{x_s} \right)}{\hat{k}\left( \bm{x_s} \right)}+\omega_l(t), \\
& \Delta\left( \bm{x_s} \right) \geq \left| \frac{{\delta_s}\left( \bm{x_s} \right)}{k\left( \bm{x_s} \right)} \right|
\end{aligned}
\end{equation}
where $\Delta\left( \bm{x_s} \right)$ represents the maximum envelope of uncertainty error in sliding mode system. Moreover, a sliding mode control law is required to regulate and stabilize the mismatch system, thus, designing a quadratic Lyapunov function formed by $V_s(s) = s^2/2$, the mismatch system is asymptotically stable if the control law satisfies:

\begin{equation}
\label{eq:SMC_controllaw}
u_s = \frac{- \varepsilon {\rm sgns} - \eta s - \hat{h}\left( \bm{x_s} \right)}{\hat{k}\left( \bm{x_s} \right)} +v_s.
\end{equation}
where $v_s = -\kappa_s\left( \bm{x_s} \right){\rm sgn}\left( s \right)$, $\kappa_s\left( \bm{x_s} \right) \geq \Delta\left( \bm{x_s} \right) + \kappa_{0},~\kappa_{0} \geq 0$, and the Lyapunov function derivative $\overset{.}{V_s}(s)$ can be obtained as follows:

\begin{equation}
\label{eq:SMC_Lya}
\begin{aligned}
 \overset{.}{V_s}\left( s\right) =&s{\overset{.}{s}}\\
 = &s[- \varepsilon\frac{k\left( \bm{x_s} \right)}{\hat{k}\left( \bm{x_s} \right)}{\rm sgns} - \eta\frac{k\left( \bm{x_s} \right)}{\hat{k}\left( \bm{x_s} \right)}s \\
 &+  h\left( \bm{x_s} \right) - \hat{h}\left( \bm{x_s} \right)\frac{k\left( \bm{x_s} \right)}{\hat{k}\left( \bm{x_s} \right)} + \omega(t) + k\left( \bm{x_s} \right)v_s]\\
 =& - \varepsilon\frac{k\left( \bm{x_s} \right)}{\hat{k}\left( \bm{x_s} \right)}\left|s\right| - \eta\frac{k\left( \bm{x_s} \right)}{\hat{k}\left( \bm{x_s} \right)}s^2 + [{\delta_s}\left( \bm{x_s} \right) + k\left( \bm{x_s} \right)v_s]s.
 \end{aligned}
\end{equation}

According to lateral-yaw dynamic model in \eqref{eq:vehicle_model}, the system function satisfies $k(\bm{x_s}) > 0$. 
The term $[{\delta_s}\left( \bm{x_s} \right) + k\left( \bm{x_s} \right)v_s]s$ in \eqref{eq:SMC_Lya} associated with model mismatch satisfies:

\begin{equation}
\begin{aligned}
 [{\delta_s}\left( \bm{x_s} \right) + k\left( \bm{x_s} \right)v_s]s &< \left| {\delta_s}\left( \bm{x_s} \right) \right|\left| s \right| + k\left( \bm{x_s} \right)v_ss \\
 &<  [\Delta\left( \bm{x_s} \right) - \beta\left( \bm{x_s} \right)]k\left( \bm{x_s} \right)\left| s \right| \\
 &< 0.
 \end{aligned}
\end{equation}

Consequently, the Lyapunov function derivative satisfies $\overset{.}{V_s}(s)<0$ with the control law \eqref{eq:SMC_controllaw}, and the sliding mode system is asymptotically stable in the presence of model mismatch and external disturbances, which means it is capable of tracking the reference $(\beta - r)$ phase trajectory as determined by the LQR trajectory tracking controller.

In the meanwhile, by solving \eqref{eq:SMC_result}, the reference differential slip ratio of the wheels on both sides $\sigma_{l}^{\text{ref}},\sigma_{r}^{\text{ref}}$ can be calculated as:

\begin{equation}
\label{eq:SMC_result}
\left\{
\begin{aligned}
 \overset{.}{v}\left( {\sigma_{l}^{\text{ref}}, \sigma_{r}^{\text{ref}}} , \alpha^{\text{des}} \right) &= \overset{.}{v}\left( \sigma_{l}^{\text{des}}, \sigma_{r}^{\text{des}}, \alpha^{\text{des}} \right) \\
 {u_s}\left( {\sigma_{l}^{\text{ref}},\sigma_{r}^{\text{ref}}} \right) &={u_s}\left( \sigma_{l}^{\text{des}}, \sigma_{r}^{\text{des}} \right)
\end{aligned}
\right.
\end{equation}
where $\sigma^{\text {des}}$ is calculated by the LMI robust controller. The reference slip ratio ${\sigma_{l}^{\text{ref}}}$ and ${\sigma_{r}^{\text{ref}}}$ will be regulated and tracked by the BSC robust controller.

% The sliding mode controller is designed to track the $(\beta - r)$ phase trajectory which is planned by the MPC controller. The SMC controller is defined by the vehicle yaw rate $r_{\rm des}$, side slip angle $\beta_{\rm des}$  and their derivatives with time that are calculated by (20), and it can be expressed as follows: 

% \begin{equation}
% \left\{
%     \begin{aligned} 
%     s &= r - r_{\rm des} + \xi\left( \beta - \beta_{\rm des} \right)\\
%     \overset{.}{s} &= \overset{.}{r} - {\overset{.}{r}}_{\rm des} + \xi( \overset{.}{\beta} - {\overset{.}{\beta}}_{\rm des} )
%     \end{aligned} 
% \right. .
% \end{equation}

% \begin{equation}
%     \overset{.}{s} = \overset{.}{r} - {\overset{.}{r}}_{\rm des} + \xi\left( \overset{.}{\beta} - {\overset{.}{\beta}}_{\rm des} \right)
% \end{equation}

% After designing the reaching law of sliding mode controller $\dot{s}=-\varepsilon \operatorname{sgn} s-\eta s$, the vehicle side slip and yaw dynamics (14) can be derived obtained by:
% \begin{equation}
% \begin{aligned}
%     \overset{.}{s} &=\overset{.}{r}\left( {\bm{x_s},u_s} \right) + \xi\overset{.}{\beta}\left( {\bm{x_s},u_s} \right) - ( {\overset{.}{r}}_{\rm des} + \xi{\overset{.}{\beta}}_{\rm des} )\\
%     &= h\left( \bm{x_s} \right) + k\left( \bm{x_s} \right)u_s.
% \end{aligned}
% \end{equation}
% where $h\left( \bm{x_s} \right) = f_1\left( \bm{x_s} \right) + \xi f_2\left( \bm{x_s} \right)$ and$k\left( \bm{x_s} \right) = g_1\left( \bm{x_s} \right) + \xi g_2\left( \bm{x_s} \right)$. 

\subsection{BSC-based Wheel System Robust Controller Design}\label{AA}

Since the wheel dynamic model is related to the moment of inertia and damping coefficient of tire, a parameter-dependent wheel system model \eqref{eq:wheel_system} is reformulated as follows:

\begin{equation}
    \begin{aligned} 
        ~{\overset{.}{x}}_{\omega}(t) &= g\left( {x}_{\omega}(t), {\theta_\omega} \right) + u_{\omega}(t) + \omega_t(t).
    \end{aligned} 
\end{equation}
where $x_{\omega} = \omega_{ij}$ and $u_{\omega} = T_{ij}$ represent the system states and control input in tire system, respectively; $\theta_\omega = [J_\omega;B_e]$ represents the parameters of wheel system; $\omega_t(t) = T_f$ denotes the disturbance caused by frictional resistance moment in a compact set $\bm{\Omega_t} \subseteq \mathbb{R}$.

For the purpose of rectifying parameter identification errors and mitigating the influence of disturbances caused by frictional resistance moments, an error system is established to align with the reference wheel angular speed $\omega_{\omega}^{\text{ref}}$ as follows:

\begin{equation}
    \begin{aligned} 
        ~{\overset{.}{e}}_{\omega}(t) &= g_e\left( e_{\omega}(t), \bm{\theta_\omega} \right) + u_{\omega}(t) + \omega_t(t).
    \end{aligned} 
\end{equation}
where $e_{\omega}=\omega_{\omega}-\omega_{\omega}^{\text{ref}}$, and $\omega_{\omega}^{\text{ref}}$ can be computed utilizing equation \eqref{eq:slip_function} in accordance with the reference slip ratio $\sigma_l^{\text{ref}}$ and $\sigma_r^{\text{ref}}$. Furthermore a nominal model based on nominal parameters $\hat{\bm{\theta}}_{\bm{\omega}}$ and excluding disturbances is reformulated as follows:

\begin{equation}
{\overset{.}{e}}_{\omega}(t) = \hat{g}_e\left( e_{\omega}(t), \hat{\bm{\theta}}_{\bm{\omega}} \right)(t) + u_{\omega}(t)
\end{equation}
where $\hat{g}_e$ represent the nominal system function. Meanwhile the differences between the actual and nominal system in bounded form satisfy:

\begin{equation}
\label{eq:bsc_error}
    \left. g_e\left( e_{\omega} \right) - \hat{g}_e\left( e_{\omega} \right) + \omega_t(t) = G\left( e_{\omega} \right); \middle| G\left( e_{\omega} \right) \middle| \leq \varrho\left( e_{\omega} \right) \right. .
\end{equation}
where $\varrho\left( e_{\omega} \right)$ represents the maximum envelope of uncertainty error in wheel error system. Furthermore, a control law is required to regulate and stabilize the mismatch system, designing a quadratic Lyapunov function formed by $V_{\omega} \left( e_{\omega} \right) = {e_{\omega}}^{2}/2$, the wheel error system is asymptotically stable if the control law satisfies:

\begin{equation}
u_{\omega} = - k_{\omega}e_{\omega} - \hat{g}_e\left( e_{\omega} \right) + v_{\omega}
\end{equation}
where $v_{\omega} = - \Gamma\left( e_{\omega} \right){\rm sgn}\left( e_{\omega} \right)$, $~\Gamma\left( e_{\omega} \right) \geq \varrho\left( e_{\omega} \right) + \Gamma_{0}$, $~\Gamma_{0} \geq 0$, and the Lyapunov function derivative $\dot{V}_{\omega}\left(e_{\omega}\right)$ can be obtained as follows:

\begin{equation}
\begin{aligned}
\dot{V}_{\omega}\left(e_{\omega}\right) = & ~e_{\omega} \dot{e}_{\omega}=e_{\omega}\left({g}_e\left(e_{\omega}\right)+u_{\omega}\right)\\
=&-k_{\omega} e_{\omega}{ }^{2}+G\left(e_{\omega}\right) e_{\omega}-\Gamma\left(e_{\omega}\right)\left|e_{\omega}\right| \\
\quad \leq&-k_{\omega} e_{\omega}{ }^{2}+\left|G\left(e_{\omega}\right)\right|\left|e_{\omega}\right|-\Gamma\left(e_{\omega}\right)\left|e_{\omega}\right| \\
=&-k_{\omega} e_{\omega}{ }^{2}+\left[\left|G\left(e_{\omega}\right)\right|-\Gamma\left(e_{\omega}\right)\right]\left|e_{\omega}\right| \leq 0.
\end{aligned}
\end{equation}

Consequently, the derivative of the Lyapunov function satisfies  $\dot{V}_{\omega}\left(e_{\omega}\right) < 0$, indicating asymptotic stability of the wheel angular speed tracking system in the presence of parameter identification errors and disturbances from frictional resistance moments.

\section{Parameter Adaptive Strategy Design}

% \textcolor{red}{To deal with the robust factors in tracking model, the $H_{\infty}$ robust controller is designed to consider the parameter uncertainties in lateral-yaw-roll dynamic, unmodeled subsystem as well as ground mechanics and aerodynamics disturbance comprehensively. Furthermore, a LPV polyhedral structure is adopted in $H_{\infty}$ robust controller, which addresses the time-varying parameters to adapt to longitudinal velocity variations, while the boundary of various robust factors is managed by novel Gaussian Process Regression model, which reduces the conservativeness. }

In this section, parameter adaptive strategies have been designed in this control framework. A RLS identification is adopted to addresses the uncertainty in time-varying parameters to adapt to tire stiffness variations; while the boundaries of various robust factors are managed by GPR model, enhancing the model accuracy and robustness of the controller.
Furthermore, the range  of uncertain parameters and the boundaries of robust factors are adjusted by Bayesian optimization, which reduces the conservatism of controller.

\subsection{Recursive Least Squares Algorithm for Online Parameter Identification}

To identify the tire parameter in vehicle longitudinal-lateral-yaw dynamics, a recognition model has been established as follows:
\begin{equation}
\bm{y_k} = \bm{\varphi_k} \cdot \bm{\theta_k} + \bm{\xi_k}
\end{equation}
where the vector $\bm{y_k} = [\dot{v}_{x},\dot{\beta}_{z},\dot{\omega}_{z}]^T$ denotes the measurement values of the actual time step $k$; the regression matrix $\bm{\varphi_k}$ denotes the historical measured values related to vehicle motion state and known parameters before the actual time step $k$; $\bm{\theta_k}$ denotes the parameter vector to be identified; $\bm{\xi_k}$ represents the present random noise in the measurement. 
For minimizing the identification error $\bm{\varepsilon_k}$, which is formulated as $\bm{\varepsilon_k} = \bm{y_k} - \bm{\varphi_k}\bm{\hat{\theta}_k}$, the parameter vector $\bm{\hat{\theta}_k}$ can be optimized in the following identification equation:

\begin{equation}
\label{eq:inversion_operation}
\begin{aligned}
\bm{\hat{\theta}_k} = \left({\bm{\varphi_k}}^T {\bm{\varphi_k}}\right)^{-1}{\bm{\varphi_k}}^T \bm{y_k}
% \hat{\boldsymbol{\theta}}[k]&=\boldsymbol{P}[k] \boldsymbol{\Phi}^T[k] \boldsymbol{y}[k]\\
% \boldsymbol{P}[k]&=\left(\boldsymbol{\Phi}^T[k] \boldsymbol{\Phi}[k]\right)^{-1}\\
\end{aligned}
\end{equation}

It is a necessary requirement that the matrix $\bm{P} = \bm{{\varphi_k}}^T \bm{{\varphi_k}}$ is invertible, while this condition is satisfied when $\bm{P}$ is positive definite or of full rank. However, as the dimensionality of matrix $\bm{{\varphi_k}}$ augments with the additional measurement signals, the inversion operation in Eq. \eqref{eq:inversion_operation} becomes more complicated, degrading the quality of parameter identification.

By iteratively refining the parameter identification without retaining the entire historical data, a Recursive Least Squares (RLS) algorithm with forgetting factor is proposed to mitigate the aforementioned challenges.
Moreover, this algorithm adaptively reduces the influence of older data, enhancing the algorithm ability to identify time-varying parameters with greater acuity, which is summarized in the following equations:

% The Forgetting Factor Recursive Least Squares (FF-RLS) algorithm, characterized by its recursive update equations with a forgetting factor, efficiently mitigates the aforementioned challenges by iteratively refining the parameter identification without retaining the entire historical data. Moreover, the algorithm adaptively reduces the influence of older data, enhancing the algorithm ability to identify time-varying parameters with greater acuity.
% The FF-RLS identification process is summarized in the following equations:

\begin{equation}
\begin{aligned}
\bm{K_k} & =\bm{P_{k-1}} \bm{\varphi_k}^T /\left(\lambda + \bm{\varphi_k} \bm{P_{k-1}} \bm{\varphi_k}^T\right) \\
\bm{\theta_k} & =\bm{\theta_{k-1}}+\bm{K_k}\left(\bm{y_k}-\bm{\varphi_k} \bm{\theta_{k-1}}\right) \\
\bm{P_k} & =\left(\bm{I}-\bm{K_k} \bm{\varphi_k}\right) \bm{P_{k-1}} / \lambda
\end{aligned}
\end{equation}
where $\bm{{K}_k}$ denotes the update gain for the identified parameter $\bm{\theta}$, while $\bm{{P}_k}$ signifies the error covariance matrix. The coefficient $\lambda$, defined in the range of $0 < \lambda < 1$, denotes the forgetting factor, determining the influence of historical data on the current parameter identification. A higher $\lambda$ is beneficial for system stability and convergence speed, while a lower $\lambda$ prioritizes recent data, improving the tracking performance of time-varying capability but increasing noise sensitivity. The following equations describe an adaptive forgetting factor method utilized to modulate the forgetting factor response to the identification error, effectively reconciling the trade-off between system stability and tracking capability:

% \begin{equation}
% \begin{aligned}
% \lambda_k & =\lambda_{\min }+\left(1-\lambda_{\min }\right) h^{\lfloor ({\frac{\varepsilon}{\sigma_\varepsilon}})^2 \rfloor} \\
% L_k & =\operatorname{NINT}\left(\rho \varepsilon_k^2\right)
% \end{aligned}
% \end{equation}

\begin{equation}
\label{eq:lammda}
\begin{aligned}
\lambda_k &= \lambda_{\min} + (1 - \lambda_{\min}) h^{q_k} \\
q_k & ={\left\lfloor ({\varepsilon_k}/{\sigma_\varepsilon})^2 \right\rfloor}
\end{aligned}
\end{equation}
where $\lfloor x \rfloor$ denotes the floor function of $x$, $\lambda_{\text{min}}$ denotes the minimum value of the forgetting factor, $h$ is a coefficient ranging from 0 to 1, ${\sigma_\varepsilon}$ denotes threshold for identification error related to sensitivity. Eq. \eqref{eq:lammda} modulates the forgetting factor within the range of $[\lambda_{\text{min}}, 1]$.

% \textcolor{red}{For the associated derivation, we refer to \cite{40}.}

% where $K_k$ is the update gain of $\theta$, and $P_k$ is the covariance of errors. The coefficient $\lambda$ is called a forgetting factor with the value of $0 < \lambda < 1$. The value of the forgetting factor is important for the identification performance, a higher value provides better convergence speed and stability while a lower value can improve the tracking capability \cite{15}. However, when the system excitation is poor, a constant forgetting factor may lead to the exponential growth of $P_k$ and make the algorithm sensitive to noise. The following equations represent a variable forgetting factor (VFF) method that can overcome the tradeoff between stability and tracking capability by adaptively changing the value of the forgetting factor \cite{14}, \cite{16}, \cite{35}:

% \begin{algorithm}
%   \caption{Euclid's algorithm}
%   \KwData{Two nonnegative integers $a$ and $b$}
%   \KwResult{$\gcd(a,b)$}
%   \eIf{$b=0$}{
%     return $a$\;
%   }{
%     return $\text{gcd}(b, a\mod b)$\;
%   }
% \end{algorithm}

\subsection{Gaussian Process Regression for Vehicle Dynamic Prediction}

Considering model mismatch caused by unmodeled subsystem and the disturbances brought by the external factors, a GPR non-parametric methodology is proposed to address aforementioned robust problems. The specifics are as follows.

In designing the robust controllers, it is determined that the robust boundary of the model mismatch and the external disturbances, i.e., the parameters of $\Delta\left( \bm{x_s} \right) $ in Eq. \eqref{eq:smc_error} and $G_e\left( e_{\omega} \right)$ in Eq. \eqref{eq:bsc_error}. Specifically, the dynamics of the autonomous vehicle is characterized by the GPR model to learn the mapping relationship from $x_s,e_{\omega}$ to $\dot{x}_s, \dot{e}_{\omega}$.
Furthermore, $\Delta\left( \bm{x_s} \right) $ and $G_e\left( e_{\omega} \right)$ will be calculated by $\dot{x}_s^{GPR}, \dot{e}_{\omega}^{GPR}$ and $\dot{x}_s^{RLS}, \dot{e}_{\omega}^{RLS}$ output from the trained GPR model and the dynamic model with identified parameters, respectively. The detailed implementation of the GPR model is given as follows.

% \textcolor{red}{Considering the model mismatch caused by uncertain parameters in the system modeling, and the unknown disturbances brought by the external factors as well as the un-modeled components of the system, a novel GPR non-parametric method is proposed in this section to solve above challenges \cite{`}. The specific details are as follows.}

% \textcolor{red}{In designing the $H_{\infty}$ controller, the considered external disturbances mainly consist of two parts, where the lateral disturbance force $F_{dy}$ and the disturbance yaw moment $M_{dz}$ are unknown disturbances, while the road curvature $\dot{\varepsilon}_{d}$, belonging to the external input, is assumed to be known. Our objective is to determine the robust boundary of the unknown disturbances, i.e., the parameters in $B_{\omega}$. Specifically, the dynamics of the commercial vehicle is characterized by a GPR non-parametric model to learn the mapping relationship from $x_d$ to $\dot{x}_d$. $\Delta\dot{x}_d$ is calculated by $\dot{x}_{dGPR}$ output from the trained GPR model and $\dot{x}_{d0}$ output from the lateral-yaw-roll dynamic model in subsection II-A. Finally, the unknown disturbances will be quantified based on the expectation of $\Delta\dot{x}_d$. }

% \textcolor{red}{The detailed implementation of the GPR model is given as follows. The model formulation is first introduced in Eq. (22).}

The observed target output $y_g$ are modeled as a joint multivariate Gaussian distribution of transformed inputs $x_g$ through the signal function $f(x_g)$, which is additionally disturbed by an independent zero-mean gaussian process noise $\epsilon$.

\begin{equation}\label{eq:22}
\begin{array}{cc}
   & y_g = f(x_g)\vspace{2ex} + \epsilon;\\
   & f(x_g) \thicksim \mathcal{G P}(m_g(x_g),k_g(x_g,x_g'))\vspace{1.5ex}; 
\end{array}
\end{equation}
where $\epsilon \sim \mathcal{N}\left(0, \sigma_\epsilon^2\right)$, the Gaussian process $\mathcal{G P}$ is a distribution over functions, defined by mean function $m_g(x_g)$ and covariance function $k_g(x_g,x_g')$. The function $k_g$ is defined as the radial basis function (RBF) as Eq. \eqref{eq:RBF}, called the kernel of the Gaussian process, which is parameterized by the hyper-parameters $\sigma_f^2$ (signal variance) and $l$ (length scale) \cite{GPR}. 

\begin{equation}
\label{eq:RBF}
\begin{array}{cc}
   k_g\left(x_g, x_g^{\prime}\right)=\sigma_f^2 \exp (-\frac{\left\|x_g-x_g^{\prime}\right\|^2}{2 l^2})
\end{array}
\end{equation}

% \begin{equation}\label{eq:22}
% \begin{array}{cc}
%    & y_g = h(x_g)^T\beta_g+f(x_g)\vspace{2ex}\\
%    & f(x_g) \thicksim N(0,K(x_g,x_g'))\vspace{1.5ex}\\
%    & K(x_i,x_j) = \sigma_f^2{\rm{exp}}(-\frac{\Vert x_i-x_j\Vert^2_2}{2l^2})
% \end{array}
% \end{equation}
% where $f(x_g)$ is characterized by a covariance structure determined by the kernel function $K(x_g,x_g')$, signifying the stochastic component. $K(x_g,x_g')$ is defined as the radial basis function (RBF), which is parameterized by the hyper-parameters $\sigma$ (signal variance) and $l$ (length scale) \cite{26}. 

% where $x_g$ denotes the model input and $y_g$ denotes the observed output. $h(x_g)$ is the basis function for dimensional conversion, $\beta_g$ is the parameter vector. $f(x_g)$ represents the zero-mean Gaussian process obeying a kernel function $K(x_g,x_g')$. Referring to \cite{26}, $K(x_g,x_g')$ is defined as the radial basis function (RBF), $\sigma,l$ are relevant hyper-parameters. 

% The training set and testing set of the model are both obtained from Trucksim simulation. Moreover, the marginal log-likelihood estimation is specifically employed as the training method.

Utilizing marginal log-likelihood estimation, the hyper-parameters $\bm{\theta_g} = [{l,\sigma_f^2,{\sigma}_{\epsilon}^2}]$ are optimized to identify the parameters that maximizes the GPR model concordance with the datasets. In this study, both the training and testing datasets are sourced from the Carsim simulation software.

\begin{equation}\label{eq:22}
\hat{l}, \hat{\sigma}_f^2, \hat{\sigma}_{\epsilon}^2 = \mathop{\arg\min}\limits_{l,\sigma_f^2,{\sigma}_{\epsilon}^2} \, {\rm{-log}}\,  P(y_g\vert x_g,\bm{\theta_g})  
\end{equation}
where $\hat{l}, \hat{\sigma}_f^2, \hat{\sigma}_{\epsilon}^2$ are the optimized hyper-parameters. Based on the kernel function with $\hat{\theta}_g$, the posterior predicted output $y^*_g$, following a multivariate normal distribution, can be calculated as follows:

\begin{equation}
y^*_g = K_g\left(x^*_g, \bm{X_g}\right)\left[K_g\left(\bm{X_g}, \bm{X_g}\right)+\sigma_\epsilon^2 \mathbf{I}\right]^{-1} \bm{Y_g}
\end{equation}
where $\bm{X_g}$ and $\bm{Y_g}$ denote the input and output variables of the training dataset, respectively. $K_g\left(\bm{X_g}, \bm{X_g}^{\prime}\right)$ represents the covariance matrix with $\bm{X_g}$, in which each element is determined by applying the kernel function $k_g$ to the corresponding pair of inputs.

\begin{algorithm}
  \caption{RLS and GPR}
  \KwIn{Training dataset $\mathbf{X}$, $\mathbf{Y}$, testing dataset $\mathbf{x}_k$, $\mathbf{y}_k$ }
  \KwOut{The predicted output $\hat{\mathbf{y}}_k^{RLS}$, $\hat{\mathbf{y}}_k^{GPR}$, the identified tire parameters $\bm{\hat{\theta}_k}$}
  % Initialize the matrix $\theta\leftarrow \theta_0$, $P\leftarrow P_0$, the parameters $\lambda \leftarrow \lambda_0$ in RLS\;
  Initialize the matrix $\bm{\theta_0}, \bm{P_0}$ and parameters $\lambda$ in RLS\;
  Convert $\mathbf{x}$ in test dataset to $\bm{\varphi_k}$\;
  % : $\theta_0, P_0, \lambda$
  % todo 定义y,phi theta分别是什么
  \For{$k\leftarrow 1$ \KwTo $N$}{
    $\bm{K_k} =\bm{P_{k-1}} \bm{\varphi_k}^T /\left(\lambda + \bm{\varphi_k} \bm{P_{k-1}} \bm{\varphi_k}^T\right)$\;
    $\bm{\hat{\theta}_k} =\bm{\hat{\theta}_{k-1}}+\bm{K_k}\left(\bm{y_k}-\bm{\varphi_k} \bm{\hat{\theta}_{k-1}}\right)$\;
    $\bm{P_k} =\left(\bm{I}-\bm{K_k} \bm{\varphi_k}\right) \bm{P_{k-1}} / \lambda$\;
    $\hat{\mathbf{y}}_k^{RLS}=\hat{\bm{\varphi_k}} \bm{\theta_{k-1}}$\;
  }
  % $\hat{\theta}_k \leftarrow  {\theta}_k$\;
  Initialize the kernel function and parameters in GPR using Eq. (50), (51)\;
  % : $\theta_g = [l, \sigma_f^2 ,\sigma_\epsilon^2]$\;
  $\hat{\theta}_g = \mathop{\arg\min}   -{\rm{log}}\,  P(\mathbf{Y}\vert \mathbf{X},\theta_g)$\;
  \For{$k\leftarrow 1$ \KwTo $N$}{
  $\hat{\mathbf{y}}_k^{GPR} = K_g\left(\mathbf{x}_k, \mathbf{X}\right)\left[K_g\left(\mathbf{X}, \mathbf{X}\right)+\sigma_\epsilon^2 \mathbf{I}\right]^{-1} \mathbf{Y}$\;
  }
  return $\hat{\mathbf{y}}_k^{RLS}$, $\hat{\mathbf{y}}_k^{GPR}$, $\bm{\hat{\theta}_k}$\;

\end{algorithm}

\subsection{Bayesian Optimization of Robust Boundary Determination and Adjustment}

To reduce conservatism in the robust controllers without compromising robustness, a Bayesian optimization approach is utilized to adaptively fine-tune the robust boundary and the range of uncertain parameters, thereby optimizing the controller's comprehensive performance.

Considering the requirements for trajectory tracking and motion stability control in AVs, a comprehensive global objective function $J_{G}$ is designed to optimize the vehicle's dynamic performance as follows:

\begin{equation}   
\begin{aligned}
    J_{G}=\sum_{k=1}^{N}\left(\left\|\bm{z_e}\right\|_{\bm{W_e}}^{2} + \|\bm{a_{v}}\|_{\bm {W_a}}^{2} + \|\bm{{\varphi_{v}}}\|_{\bm {W_\varphi}}^{2}\right)
    % ({\varphi_{i}} + W_{\beta}{\dot{\beta}}^2)
\end{aligned}
\end{equation}
where 
$\bm{z_e}, \bm{a_{v}}, \bm{{\varphi_{v}}}$ denotes the vectors of tracking error, motion acceleration and tire friction utilization. ${\bm {W_e}}, {\bm {W_a}}, {\bm {W_\varphi}}$ represents the global weight matrices for tracking accuracy, control smoothness and driving stability, respectively. 

Given the correlation between system performance and the robust scaling coefficient $\alpha_b$, the global objective function is defined as $J_G = J_G(\alpha_b)$. A Bayesian optimization method based on the Upper Confidence Bound (UCB) algorithm is adopted to modify $\alpha_b$ in order to minimize the global objective value.

\begin{equation}\label{eq:23}
\begin{aligned}
&\hat{\bm{\alpha_b}} = \mathop{\arg\min}\limits_{\bm{\alpha_b}} \, J_G(\bm{\alpha_b})\\
\end{aligned}
\end{equation}
where $\bm{\alpha_b} = [\alpha_{b\theta}, \alpha_{bi}, \alpha_{be}]$, which are set within the range of $[0, +\infty]$. The iterative approach using UCB as the acquisition function is as follows:

\begin{equation}\label{eq:23}
\begin{aligned}
% \bm{\alpha_b}_{,k+1}=\arg \min \left(UCB(\bm{\alpha_b} ; {D}_{1: k})\right)\\
\bm{\alpha_{b,k+1}}=\arg \min \left({\mu}_b(\bm{\alpha_b})+\kappa_b {\sigma}_b(\bm{\alpha_b})\right)
\end{aligned}
\end{equation}
where $\kappa_b$ denotes the proportional coefficient, ${\mu}_b$ and ${\sigma}_b$ denote the mean function and standard deviation function of $J_G$. In the application, an iterative algorithm based on Bayesian optimization is employed to identify the optimal parameters. The pseudocode for this algorithm is as follows:

\begin{algorithm}
  \caption{Robust Boundary Determination and Optimization}
  \KwIn{ $\mathbf{y}_k$, $\hat{\mathbf{y}}_k^{RLS}$, $\hat{\mathbf{y}}_k^{GPR}$, $\bm{\hat{\theta}_k}$ in Algorithm 1}
  \KwOut{Robust boundary $\Theta, \bm{\Omega_s}, \bm{\Omega_t}$, their scaling coefficient $\hat{\alpha}_b$}
  $e_{ik} \leftarrow \hat{\mathbf{y}}_k^{GPR} - \hat{\mathbf{y}}_k^{RLS}$\;
  $e_{ek} \leftarrow {\mathbf{y}}_k - \hat{\mathbf{y}}_k^{GPR}$\;
  Initialize iteration times $N_{\alpha}$\;
  $\{\Theta, \Upsilon\} \leftarrow\left\{\bm{{\alpha}_{b0}}, J_G(\bm{{\alpha}_{b0}})\right\}$\;
  \For{$i\leftarrow 1$ \KwTo $N_{\alpha}$}{
    $\bm{{\alpha}_b}^i\leftarrow\arg \max \operatorname{UCB}(\Theta \mid\{\Theta, \Upsilon\})$\;
    $\hat{\theta}_k^{max} \leftarrow \hat{\theta}_k^{max}{\alpha}_{b\theta}^i$, $\hat{\theta}_k^{min} \leftarrow \hat{\theta}_k^{min}/{\alpha}_{b\theta}^i$\;
    $e_{ik}^{max} \leftarrow e_{ik}^{max}{\alpha}_{bi}^i$, $e_{ik}^{min} \leftarrow e_{ik}^{min}/{\alpha}_{bi}^i$\;
    $e_{ek}^{max} \leftarrow e_{ek}^{max}{\alpha}_{be}^i$, $e_{ek}^{min} \leftarrow e_{ek}^{min}/{\alpha}_{be}^i$\;
    Find a polytope $\forall  \hat\theta_k \in \Theta \subseteq \mathbb{R}^{n_\theta}$\;
    Find compact sets $\bm{\Omega_s} \subseteq \mathbb{R}^{n_\omega}$ and $\bm{\Omega_t} \subseteq \mathbb{R}$\;
    $J_G^i(\bm{{\alpha}_b}^i)=\text { CalculateGlobalCost } (\bm{{\alpha}_b}^i, \Theta, \bm{\Omega_s}, \bm{\Omega_t})$\;
    $\{\Theta, \Upsilon\} \leftarrow\{\Theta, \Upsilon\} \cup\left\{\bm{{\alpha}_b}^i, J_G(\bm{{\alpha}_b}^i)\right\}$\;
    $\bm{\hat{\alpha}_b}=\arg \min J_G(\bm{{\alpha}_b})$\;
  }
  return $\bm{\hat{\alpha}_b}, \Theta, \bm{\Omega_s}, \bm{\Omega_t}$\;
  \textit{function} CalculateGlobalCost($\bm{\hat{\alpha}_b}^i, \Theta, \bm{\Omega_s}, \bm{\Omega_t}$)\;
  ~~Take $\Theta, \bm{\Omega_s}, \bm{\Omega_t}$ into Eq. (23), (30), (39)\;
  ~~Carry out simulation experiment and record data\;
  ~~$J_{G}^i=\sum_{k=1}^{N}\left(\left\|\bm{z_e}\right\|_{\bm{W_e}}^{2} + \|\bm{a_{v}}\|_{\bm {W_a}}^{2} + \|\bm{{\varphi_{v}}}\|_{\bm {W_\varphi}}^{2}\right)$\;
  return $J_{G}^i$

\end{algorithm}

\begin{figure}[h]
    \captionsetup{font={scriptsize}}
    \centerline{\includegraphics[width=0.35\textwidth]{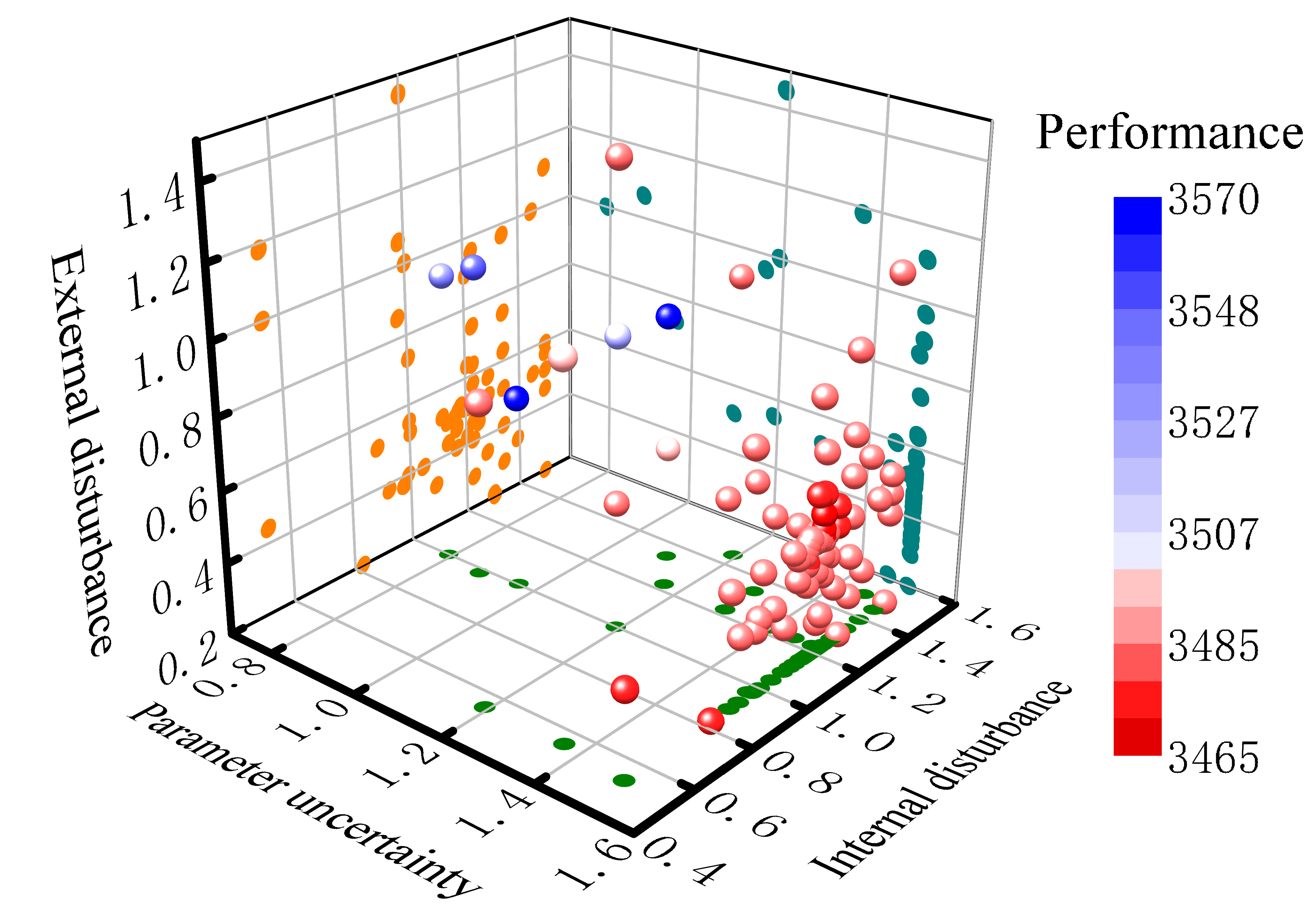}}
    \caption{The adjusted scaling coefficients of robust boundary by Bayesian optimization.}
    \label{fig}
\end{figure}

Through algorithm 2, the robust scaling coefficients have been determined, which are shown in Fig. 3. The scaling coefficients of parameter uncertainty, internal disturbance and external disturbance are determined as 1.50, 1.22 and 0.51 respectively, which will be adopted to  simulation in section V. Furthermore, this results also elucidate that concerning parameter uncertainty and internal disturbances, the boundaries of robust factors need to be enlarged to enhance the system's robustness. For external disturbances, the boundaries of robust factors should be narrowed to decrease the conservatism of system. 

\section{Experiment Validation}

% \textcolor{red}{In this section, the performance of the proposed gain-scheduled LPV/$H_{\infty}$ controller for commercial vehicles will be evaluated via Trucksim-Simulink co-simulation and hardware-in-loop (HIL) experiments. The proposed control strategy will be compared with three standard baseline controllers in the classic tracking scenario to verify its effectiveness.}

In this section, the performance of the proposed parameter adaptive framework for autonomous vehicles will be evaluated through MATLAB/Simulink and Carsim joint simulation experiments. The proposed control methodology will be compared with two baseline controllers in the extreme tracking scenario to validate its effectiveness.

% \subsection{Simulation with Analysis}

To verify the effectiveness and performance of the proposed scheme, the MATLAB/Simulink and Carsim joint simulation is built. 
The dynamic model of E class hatchback and road are designed in Carsim, which is performed through the simfile interface in MATLAB software.
The parameters of the vehicle used in Carsim and for controller design are enumerated in Table I.

% \begin{table}[htbp]
% \centering
% \rowcolors{2}{gray!10}{white} % 为表格的奇数行添加淡灰色背景
% \begin{tabular}{ccc}
% \toprule
% Symbol & Parameter Description & Value \\
% \midrule
% ms & Sprung mass & 1989 kg \\
% Ir & Vehicle roll inertia & 649 kg \cdot m^2 \\
% Iz & Vehicle yaw inertia & 4650 kg \cdot m^2 \\
% Tu & Track width & 1.6 m \\
% a & CG to front axle distance & 1.43 m \\
% b & CG to rear axle distance & 1.42 m \\
% hs & CG roll height & 0.72 m \\
% mu & Unsprung mass & 224 kg \\
% Iw & Wheel spin inertia & 1.7 kg \cdot m^2 \\
% Ks & Suspension roll stiffness & 146000 N/m \\
% Cs & Suspension roll damping & 10500 N \cdot s/m \\
% Ri & Tire radius & 0.34 m \\
% UQi,mas & Max motor/brake torque & 1600 N \cdot m \\
% \bottomrule
% \end{tabular}
% \caption{Vehicle Parameters}
% \end{table}

% \begin{table}[h]
%     \centering
%     \captionsetup{font={scriptsize}}
%     \fontsize{8pt}{\baselineskip}\selectfont{
%     \caption{Vehicle parameters of the controller  }
%     \begin{tabular}{ccc}\hline
%         % Setting&\multicolumn{2}{c|}{A4 size paper}\\\hline
%         Parameter & Name &  Value \\ \hline
%         m & Vehicle mass & 1653 kg  \\
%         $I_{z}$ & Yaw moment of inertia  & 3234 $\rm{kg} \cdot \rm{m}^{2}$  \\
%         $l_{f}$ & Front axle distances to mass center & 1.402 m  \\
%         $l_{r}$ & Rear axle distances to mass center & 1.646 m  \\
%         $C_{\alpha}$ & Lateral stiffness coefficient  & 64934.5  \\
%         $C_{\sigma}$ & Longitudinal stiffness coefficient & 63292.5 \\\hline
%     \end{tabular}
%     \label{tab:Margin_settings}
%     }
% \end{table}

\begin{table}[h]
    \centering
    \captionsetup{font=scriptsize}
    \fontsize{8pt}{\baselineskip}\selectfont
    \caption{Vehicle parameters of the controller}
    \rowcolors{2}{white}{gray!10} % 从第二行开始，每隔一行使用淡灰色背景
    \begin{tabular}{ccc}
        \toprule
        Parameter & Name &  Value \\ \hline
        $m$ & Vehicle mass & 1653 kg  \\
        $I_z$ & Yaw moment of inertia & $3234 \, \text{kg} \cdot \text{m}^2$  \\
        $l_f$ & Front axle distances to mass center & 1.402 m  \\
        $l_r$ & Rear axle distances to mass center & 1.646 m  \\
        $h$ & Height of CG of sprung mass from roll axis & 0.57 m  \\
        $C_\alpha$ & Lateral stiffness coefficient  & 64934.5  \\
        $C_\sigma$ & Longitudinal stiffness coefficient & 63292.5 \\ \bottomrule
    \end{tabular}
    \label{tab:Margin_settings}
\end{table}

The framework of experiment is displayed in Fig. 4, in which two baseline controllers MPC \cite{duibi_MPC} and LMI \cite{duibi_LMI} are adopted for comparison with the proposed parameter adaptive robust control with LQR framework (ARC).
It must be stated that ARC enhances its performance by providing a more precise identification of robust boundary parameters, while the LMI controller adopts more conservative fixed boundary parameters.

\begin{figure}[htbp]
    \centering
    \includegraphics[width=3.4in]{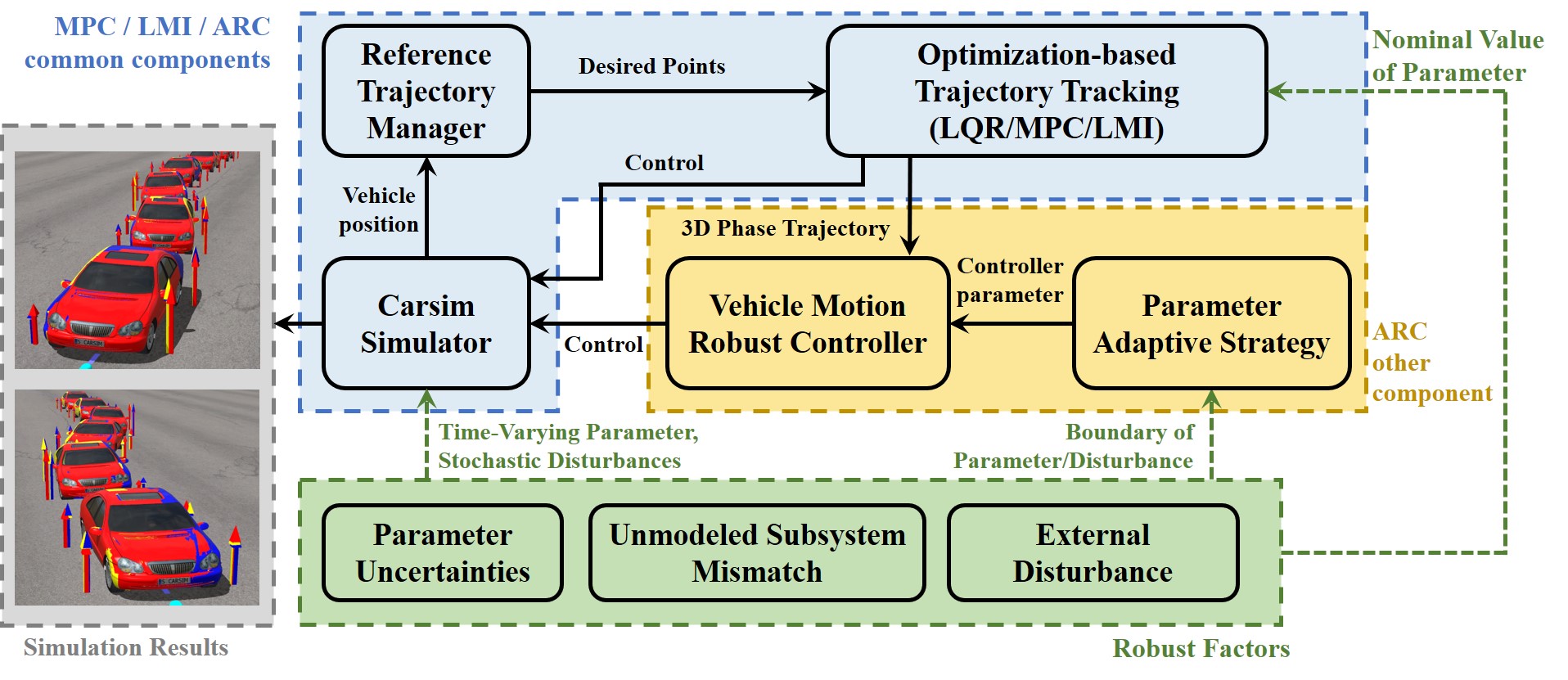}\\
    \caption{An overview of the comparative simulation framework and its relationship with robust factors and system components.}
    \label{fig.1l}
\end{figure}

A double lane change scenario with a high reference speed, representing emergency obstacle avoidance, is adopted in experiment to test the controller performances under extreme condition, which is challenging to accomplish within the context of actual driving.
In this case, reference trajectory is depicted in Fig. 5, while the vehicle longitudinal velocity is fixed and set to 60km/h, investigating the effect of the external disturbance in longitudinal dynamic quantitatively, such as tire rolling resistance moment.

In this simulation, three robust factors including uncertain time-varying parameters, unmodeled subsystems, and external disturbances are set to comprehensively evaluate the robust performance of the controller.
Due to different driving conditions, changes in the vertical force and tire loading will lead to bounded time-varying uncertainties in the lateral and longitudinal stiffness coefficient of the tires.
In the meanwhile, relative to the model in controller, there naturally exist unmodeled subsystems in the multi-dimensional high-dynamic CarSim model affecting the vehicle's motion, such as aerodynamic mechanics and suspension systems, which can be regarded as unknown but bounded model mismatches.
Furthermore, random disturbances of lateral force and yaw disturbance with extreme values of 1000 N and 1000 N.m, respectively, are artificially introduced to investigate the impact of external disturbances on the controller's performance.

Figs. 5(a)-(c) demonstrate the superiority of the proposed ARC controller performance in comparison with the two baseline controllers from different aspects.
As shown in Fig. 5(a), these three controllers can successfully achieve trajectory tracking, but the ARC controller achieves higher tracking accuracy.
Moreover, the maximum tracking error of ARC (0.043m) significantly surpasses that of the other controllers (0.113m and 0.104m, respectively), further demonstrating the superior tracking accuracy of ARC, shown in Fig. 5(b).
Meanwhile, Fig. 5(c) visualizes the dynamic behavior of the steering angle, while the ARC controller generates smoother and more reasonable steering angle with the least oscillation, under the influence of various robust factors.

% Hence, in comparison with the MPC and LMI controllers, the control framework of ARC more effectively achieves a dual enhancement in tracking accuracy and mitigation of control oscillations. Compared with the MRC controller, through parameter identification and optimization techniques to ascertain and adjust the boundaries of robust factors (encompassing parametric uncertainties, unmodeled subsystems, and external disturbances), the ARC controller significantly diminishes the system's inherent conservatism, consequently ameliorating the overall control performance. 
Hence, in comparison with the MPC and LMI controllers, the control framework of ARC more effectively achieves a dual enhancement in tracking accuracy and mitigation of control oscillations. Compared with the LMI controller, through parameter identification and optimization techniques to ascertain and adjust the boundaries of robust factors (encompassing parametric uncertainties, unmodeled subsystems, and external disturbances), the ARC controller significantly diminishes the system's inherent conservatism, consequently ameliorating the overall control performance. 
Although this approach may lead to an increase in the peak values of the steering angle, it is deemed acceptable provided that such increases remain within reasonable limits and do not introduce any disadvantages pertaining to vehicle stability.

Figs. 6(a)-(b) illustrate the yaw stability during trajectory tracking process.
In comparison with the MPC and LMI controllers, the ARC controller demonstrates a smaller and smoother sideslip angle and angular velocity, indicating a substantial enhancement in yaw stability.
Furthermore, under the influence of robust factors such as external disturbances, the ARC controller also exhibits disturbance rejection capabilities, with a notable reduction in oscillations in both the sideslip angle and angular velocity, thereby affirming the superior robustness of the ARC controller.

\begin{figure}[h]
    \captionsetup{font={scriptsize}}
    \centerline{\includegraphics[width=0.48\textwidth]{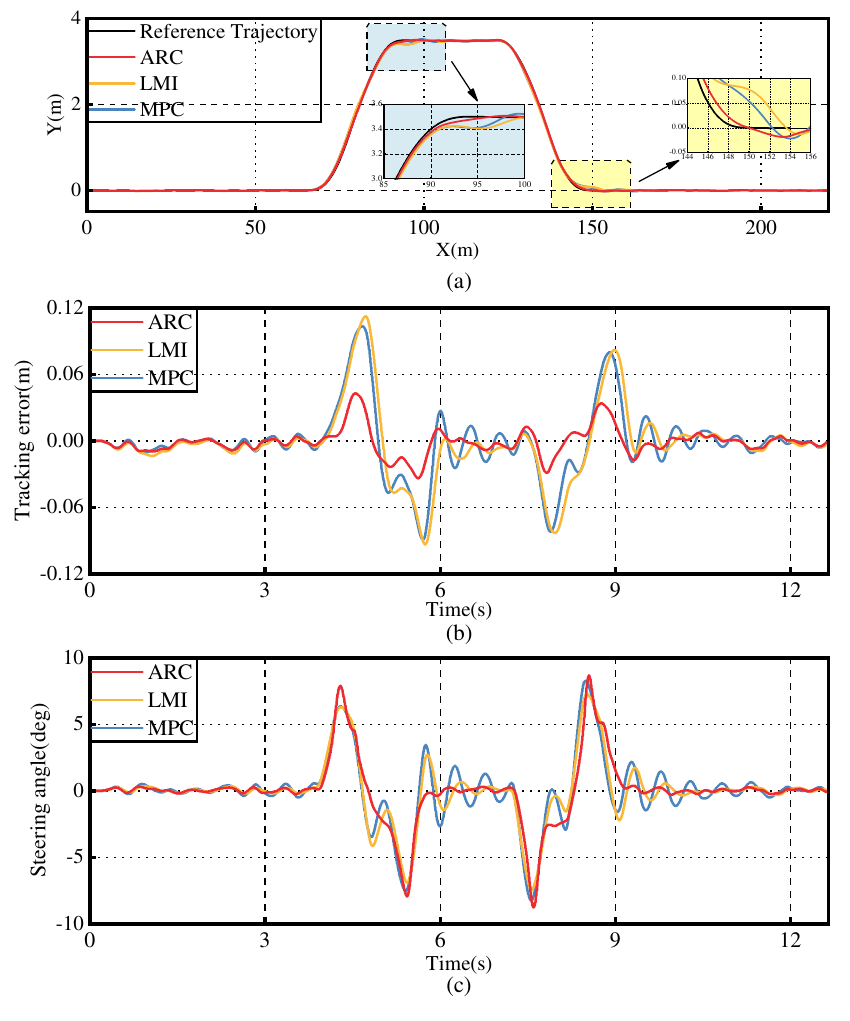}}
    \caption{The driving trajectory (a), tracking error (b) and steering angle (c) in simulation results of three controllers.}
    \label{fig}
\end{figure}

\begin{figure}[h]
    \captionsetup{font={scriptsize}}
    \centerline{\includegraphics[width=0.48\textwidth]{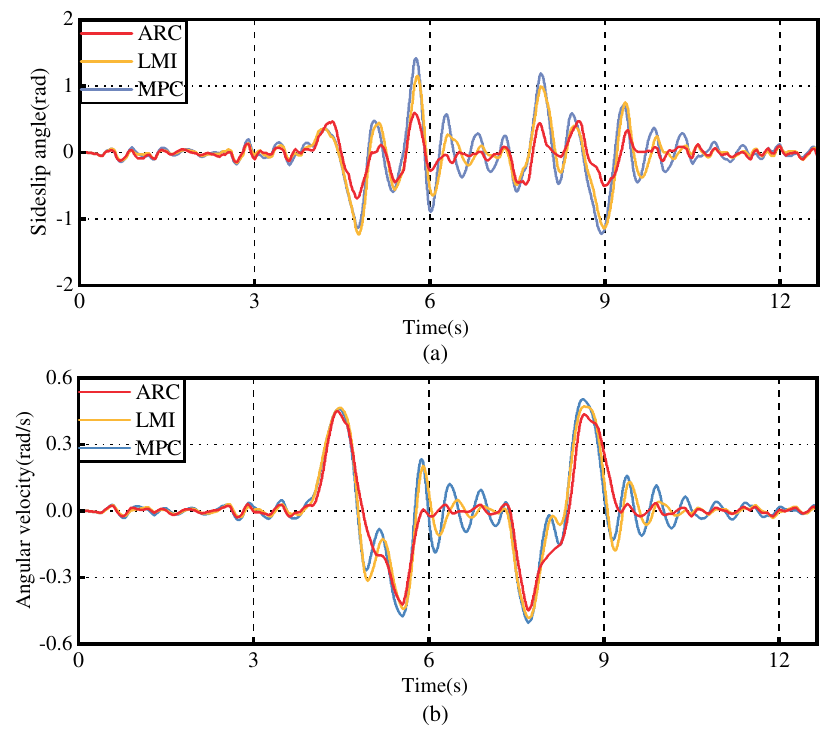}}
    \caption{The sideslip angle (a) and angular velocity (b) in simulation results of three controllers.}
    \label{fig}
\end{figure}

% Fig. 6 (b) illustrates the performance of braking control input and brake-scheduling parameter. Similar to the analysis of $\rho_{\delta_f}$, it can be inferred from the result of brake-scheduling parameter that the strategy with the GPR module shows a more reasonable scheduling of $\rho_{M_z}$ and output appropriate direct lateral yaw moments at turning locations, which significantly reduces the effect of the DYC intervention on the longitudinal dynamics. Accordingly, the multi-objective optimization of commercial vehicle's lateral control automation can be achieved through the seamless coordination of steering and braking as input.

\begin{figure}[H]
\centering
    \includegraphics[width=3.5in]{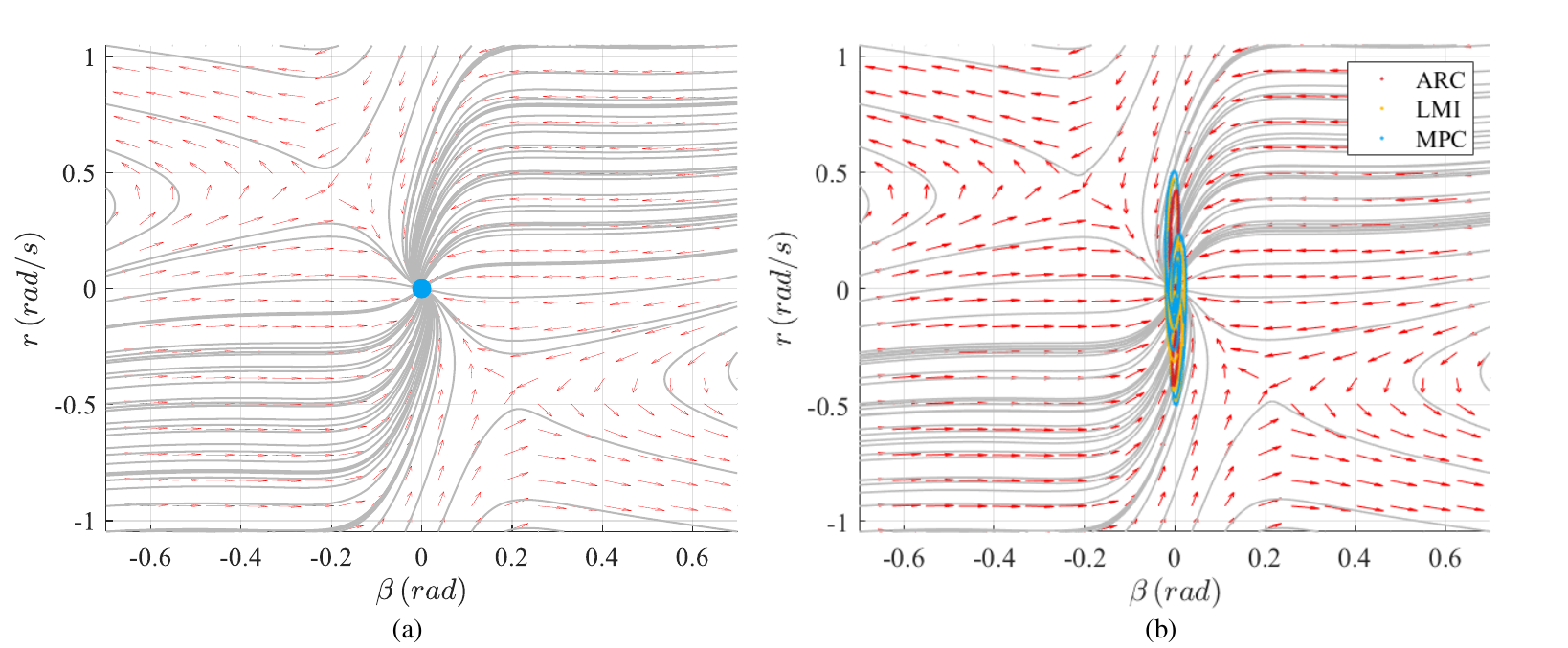}\\
\caption{The 2D phase trajectory of sideslip angle and angular velocity in the phase plane for the three controllers ARC, LMI, MPC ($V_x = 16.6\textrm{m/s}, \delta = 0\textrm{rad}$).}
\label{fig.5}
\end{figure}

% \begin{figure}[H]
% \centering
%     \includegraphics[width=3.5in]{fig//fig13.pdf}\\
% \caption{Vehicle stability analysis based on phase plane ($V_x = 18\textrm{m/s}, \delta_f = 0.088\textrm{rad}$). (a)-(d) represent the phase plane trajectory and phase trajectory points of the four controllers GS-GPR, GS, GF-GPR and GF, respectively.}
% \label{fig.5}
% \end{figure}

\begin{figure}[h]
    \captionsetup{font={scriptsize}}
    \centerline{\includegraphics[width=0.5\textwidth]{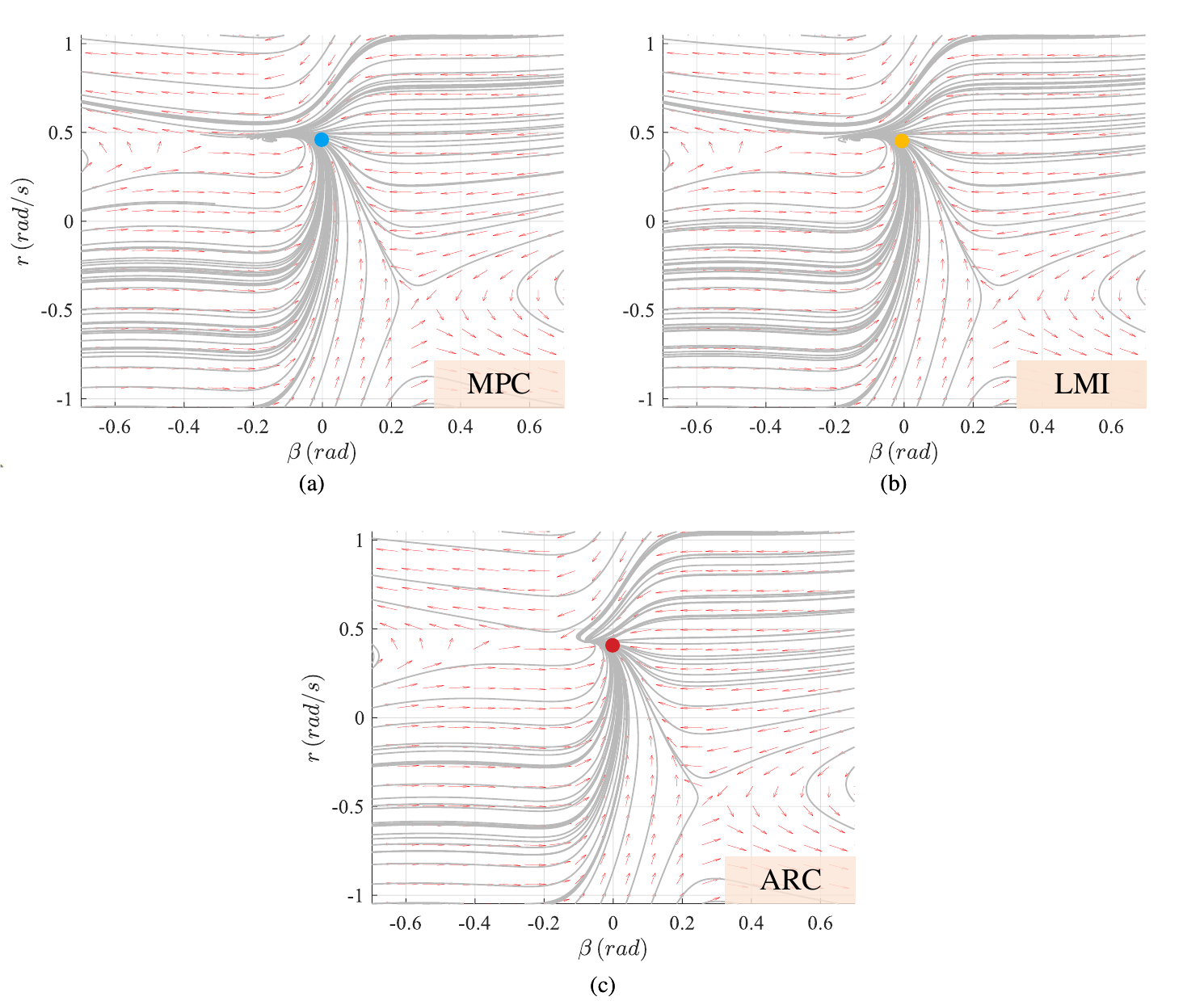}}
    \caption{$V_x = 16.51\textrm{m/s}, \delta_f = 0.079\textrm{rad}$. (a)-(c) represent the phase plane trajectory and phase trajectory points of the three controllers ARC, LMI and MPC, respectively.}
    \label{fig}
\end{figure}

\begin{figure}[h]
    \captionsetup{font={scriptsize}}
    \centerline{\includegraphics[width=0.5\textwidth]{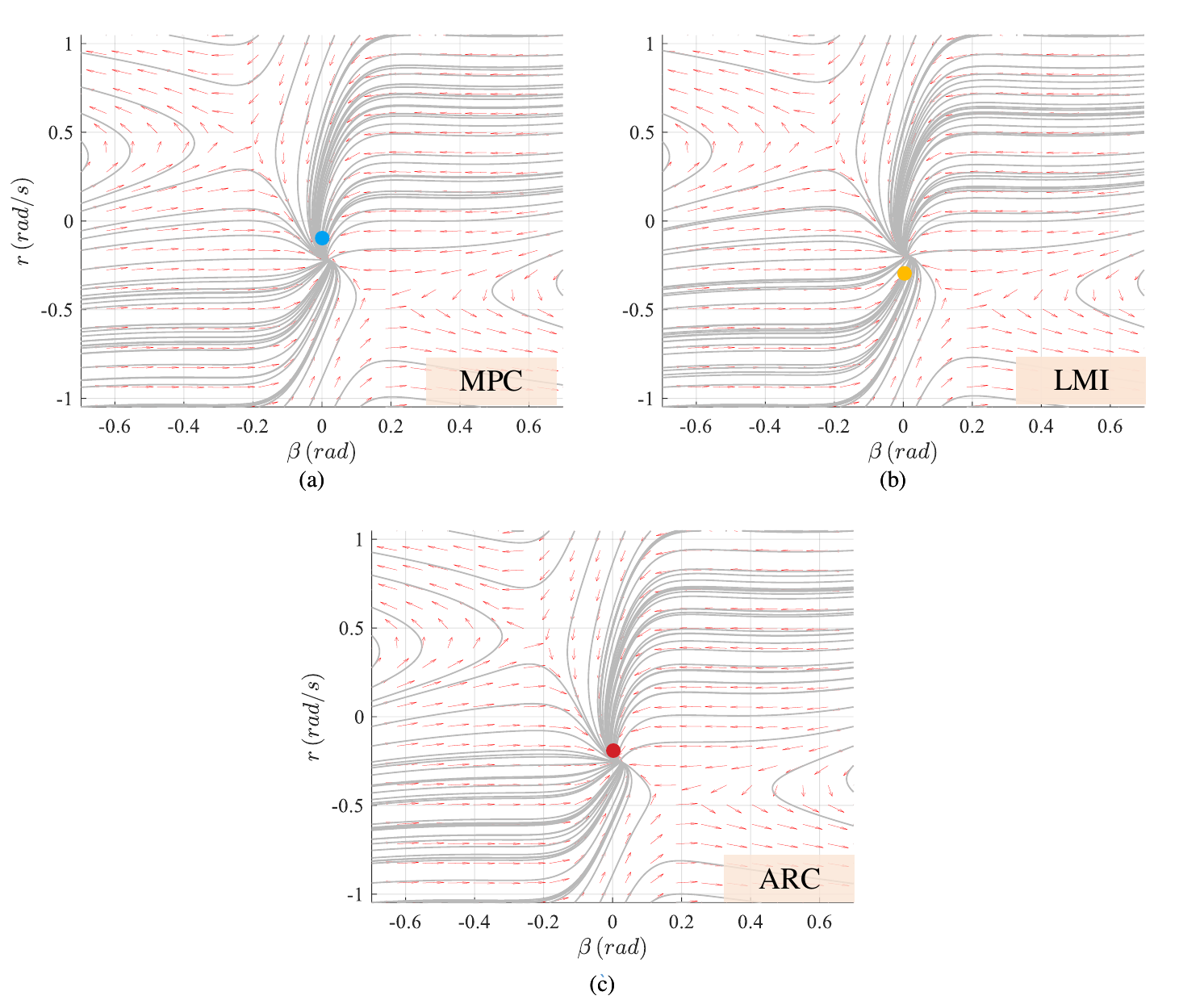}}
    \caption{$V_x = 16.54\textrm{m/s}, \delta_f = -0.046\textrm{rad}$. (a)-(c) represent the phase plane trajectory and phase trajectory points of the three controllers ARC, LMI and MPC, respectively.}
    \label{fig}
\end{figure}

% \begin{figure}[h]
%     \captionsetup{font={scriptsize}}
%     \centerline{\includegraphics[width=0.46\textwidth]{fig_new/Phase_trajectory_1.jpg}}
%     \caption{$V_x = 16.51\textrm{m/s}, \delta_f = 0.079\textrm{rad}$). (a)-(d) represent the phase plane trajectory and phase trajectory points of the four controllers GS-GPR, GS, GF-GPR and GF, respectively.}
%     \label{fig}
% \end{figure}

% \begin{figure}[h]
%     \captionsetup{font={scriptsize}}
%     \centerline{\includegraphics[width=0.46\textwidth]{fig_new/Phase_trajectory_2.jpg}}
%     \caption{$V_x = 16.54\textrm{m/s}, \delta_f = -0.046\textrm{rad}$). (a)-(d) represent the phase plane trajectory and phase trajectory points of the four controllers GS-GPR, GS, GF-GPR and GF, respectively.}
%     \label{fig}
% \end{figure}

To further analyze vehicle yaw stability, a phase plane analysis method has been utilized as shown in Fig. 7(a). This approach involves plotting the phase plane diagram with the yaw rate and sideslip angle as variables, characterizing the vehicle lateral-yaw dynamics at a specific steering angle and longitudinal velocity. Concurrently, the phase plane diagram reveals a stable equilibrium point, characterized by the progressive convergence of neighboring trajectories towards it, delineating the region known as the attraction domain.

As depicted in Fig. 7(b), the phase trajectory of the ARC controller is primarily concentrated in the vicinity of the origin, contrasting with the broader trajectory dispersion in other controllers. 
Significantly, phase trajectories that exhibit greater divergence and oscillatory behavior are indicative of a vehicle's substantial deviation from a stable equilibrium state, which in turn renders the vehicle more susceptible to destabilization.
Consequently, it substantiates the advantages of yaw stability in ARC controller framework, demonstrating insensitivity to variations in curvature and external disturbances.

Additionally, Figs. 8(a)-(c) and Figs. 9(a)-(c) showcases the phase planes of the three controllers at the similar steering angle and motion state. It is illustrated that when the projected point, representing the actual side-slip angle and yaw angular velocity (denoted by a dot), is in closer proximity to the stable equilibrium on the phase plane, the vehicle's lateral-yaw motion tends to converge towards this equilibrium, thereby ensuring system stability. Enhanced lateral and yaw stability are consequently achieved. Comparatively, the ARC's state point resides closer to the phase plane's stable equilibrium point than those of the LMI and MPC controllers, underscoring the superior yaw stability attributed to the ARC controller.

% \textcolor{red}{Consequently, the GS-GPR could improve the motion stability significantly, especially the vehicle yaw stability, through coordinating steering and braking control in different driving condition based on gain-schedule control strategy.}

Consequently, the proposed framework could improve the tracking performance and driving stability significantly, especially the yaw stability of vehicle, through the LQR and robust controllers based on parameter adaptive strategy.

\section{Conclusion} 
In this study, a parameter adaptive control framework for autonomous vehicles is proposed, which adopts linear quadratic regulator and robust control strategy.
Without introducing complexity into each controller, this control framework isolates the trajectory tracking problem from motion control, synchronously improving tracking performance and driving stability.
It also establishes three robust controllers to consider multiple robust factors, in which the uncertainty in time-varying parameters as well as the boundaries of model mismatch and external disturbance are addressed through the RLS identification and GPR model respectively, enhancing the robust performance.
The range  of uncertain parameters and the boundaries of robust factors are adjusted by Bayesian optimization, reducing the conservatism of the controller.
The advantages of the proposed control framework are verified on the MATLAB/Simulink and Carsim joint simulation platform.
The experimental results demonstrate that the proposed methodology effectively enhance tracking performance and driving stability, while determining and addressing robust factors caused by the parameter uncertainties, mismatch of unmodeled subsystem and external disturbance elevates the robust performance and reduces the conservatism.

In future work, subsequent research endeavors will concentrate on confirming the effectiveness of the proposed methodology within more intricate and extreme driving conditions.

% \textcolor{red}{On the joint simulation platform of MATLAB/Simulink and Carsim, the advantages of the proposed control framework is verified in an extreme scenario.
% The experimental results indicate that compared with the former controller, the proposed scheme effectively improves tracking accuracy and driving stability, while addressing robustness issues caused by the model mismatch, external unknown disturbances and time-varying parameters improves the robust performance and reduces the conservation significantly.}

% \begin{figure*}[t]
% \centering
% \subfloat[]{\includegraphics[width=3.5in]{fig//fig9.pdf}
% \label{a}}
% \subfloat[]{\includegraphics[width=3.5in]{fig//fig10.pdf}
% \label{b}}\\
% \subfloat[]{\includegraphics[width=3.65in]{fig//fig11.pdf}
% \label{c}}
% \subfloat[]{\includegraphics[width=3.65in]{fig//fig12.pdf}
% \label{d}}
% \caption{Simulation results of large curvature extreme maneuver. (a) shows macro tracking performance and (b) demonstrates the tracking accuracy. (c) and (d) visualize the evolution of specific indicators and scheduling parameters during lateral control automation.}
% \label{fig.1}
% \end{figure*}

\ifCLASSOPTIONcaptionsoff
  \newpage
\fi

\bibliographystyle{IEEEtran}
\bibliography{IEEEabrv,name}

\begin{IEEEbiography}[{\includegraphics[width=1in,height=1.25in,clip,keepaspectratio]{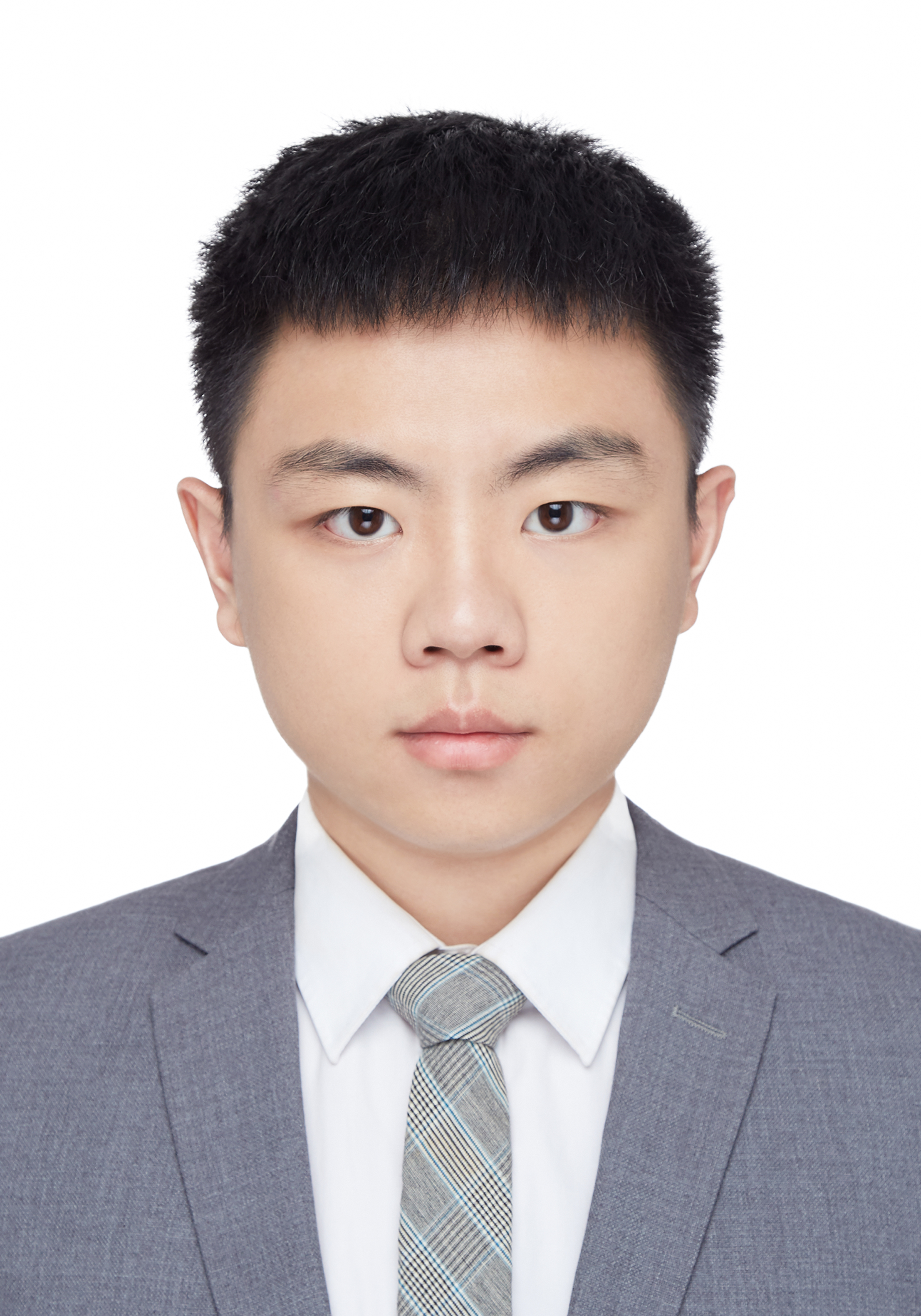}}]{Jiarui Song}
received the B.S. degree a from the Beijing Institute of Technology, Beijing, China, in 2020. He received the M.S. degree in mechanical engineering with the Beijing Institute of Technology, Beijing, China, in 2023. He is currently working towards the Ph.D. degree in mechanical engineering from School of Vehicle and Mobility, Tsinghua University, Beijing, China. 

His research interests include autonomous vehicles, intelligent control and vehicle system dynamics.
\end{IEEEbiography}

\begin{IEEEbiography}[{\includegraphics[width=1in,height=1.25in,clip,keepaspectratio]{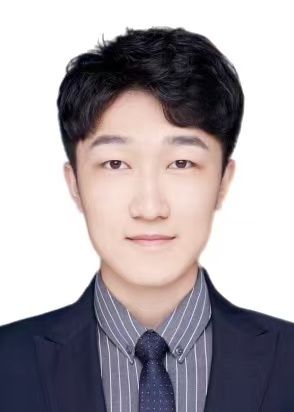}}]{Yingbo Sun}
received the B.S. degree from Tsinghua University, Beijing, China in 2020. He is currently working towards the Ph.D. degree in mechanical engineering from School of Vehicle and Mobility, Tsinghua University, Beijing, China. 

His research interests include trajectory prediction and decision making of connected and automated vehicles and unsignalized roundabout cooperation.
\end{IEEEbiography}

\begin{IEEEbiography}[{\includegraphics[width=1in,height=1.25in,clip,keepaspectratio]{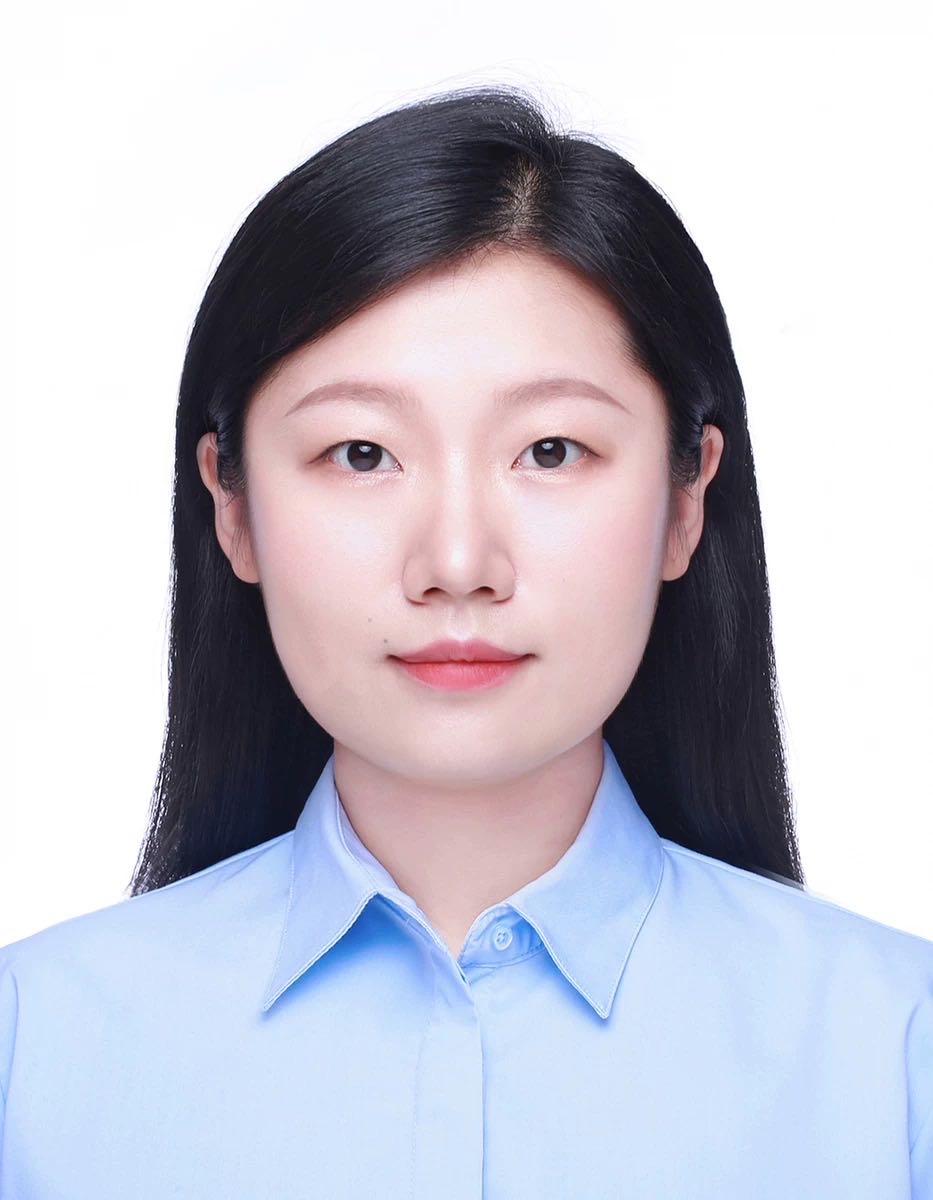}}]{Qing Dong}
received the bachelor’s degree in energy and power engineering from College of Automotive Engineering, Jilin University, Changchun, China, in 2019. She is currently working towards the Ph.D. degree in mechanical engineering from School of Vehicle and Mobility, Tsinghua University, Beijing, China. 

Her research interests include a dynamic control, autonomous driving, and intelligent transportation systems.
\end{IEEEbiography}

\begin{IEEEbiography}[{\includegraphics[width=1in,height=1.25in,clip,keepaspectratio]{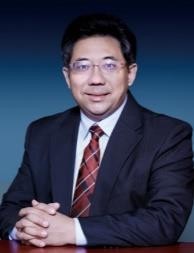}}]{Xuewu Ji}
(Member, IEEE) received his B.S., M.S. and Ph.D. degrees in automotive engineering from the College of Automotive Engineering, Jilin University, China, in 1987, 1990 and 1994, respectively. He is currently a Professor in mechanical engineering from School of Vehicle and Mobility, Tsinghua University, Beijing, China. 

His research interests include vehicle dynamics and control, advanced steering system technology, intelligent vehicle planning and decision-making, vehicle trajectory prediction, and intelligent transportation safety.

Dr. Ji received the National Science and Technology Progress Award for his achievements in the industrialization of electric power steering technology in 2014. He is the vice chairman of the Automotive Steering Technology Sub-Committee of the Chinese Society of Automotive Engineering.
\end{IEEEbiography}

\end{document}